\documentclass{aa}
\usepackage{txfonts}
\usepackage{graphicx}
\usepackage[bookmarks, breaklinks=true,
  colorlinks=true, allcolors=blue, pdftitle={Evolution of
    critically-rotating accretors in Algols},
  pdfauthor={R. Deschamps}, pdfkeywords={Binaries: general, Stars:
    rotation, Accretion, accretion disc, Stars: magnetic field, Stars:
    evolution, Methods: numerical}]{hyperref}
\usepackage{natbib}
\bibpunct{(}{)}{;}{a}{}{,} % to follow the A&A style

\begin{document}

\title{Critically-rotating accretors and non-conservative
  evolution in Algols}

\author{R. Deschamps\inst{1} \and L. Siess\inst{1} \and P.~J. Davis\inst{1}
  \and A. Jorissen\inst{1}}
\date{Received 19 March 2013 / Accepted 5 July 2013 }

\offprints{Romain.Deschamps@ulb.ac.be}

\institute{Institut d'Astronomie et d'Astrophysique, Universit\'e
  Libre de Bruxelles, ULB, CP 226, 1050 Brussels, Belgium}

\abstract {During the mass-transfer phase in Algol systems, a large amount
  of mass and angular momentum are accreted by the gainer star which can be
  accelerated up to its critical Keplerian velocity. The fate of the gainer
  once it reaches this critical value is unclear.}%context
{We investigate the orbital and stellar spin evolution in semi-detached
  binary systems, specifically for systems with rapidly rotating
  accretors. Our aim is to better distinguish between the different
  spin-down mechanisms proposed which can consistently explain the slow
  rotation observed in Algols' final states and assess the degree of
  non-conservatism due to the formation of a hotspot.} %aims
{We use our state-of-the-art binary evolution code, \textsc{Binstar}, which
  incorporates a detailed treatment of the orbital and stellar spin,
  including all torques due to mass transfer, the interactions between a
  star and its accretion disc, tidal effects and magnetic braking. We also
  present a new prescription for mass loss due to the formation of a
  hotspot based on energy conservation.} %methods
{The coupling between the star and the disc via the boundary layer prevents
  the gainer from exceeding the critical rotation. Magnetic-field effects,
  although operating, are not the dominant spin-down mechanism for sensible
  field strengths. Spin down owing to tides is 2--4 orders of magnitudes
  too weak to compensate the spinning-up torque due to mass
  accretion. Moreover, we find that the final separation strongly depends
  on the spin-down mechanism.  The formation of a hotspot leads to a large
  event of mass loss during the rapid phase of mass transfer. The degree of
  conservatism strongly depends on the opacity of the impacted
  material.} %results
{A statistical study as well as new observational constraints are needed to
  find the optimal set of parameters (magnetic-field strength, hotspot
  geometry,...) to reproduce Algol evolutions.}%conclusion

\keywords {Binaries: general -- Stars: rotation -- Accretion,
  accretion disks -- Stars: magnetic field -- Stars: evolution --
  Methods: numerical}

\maketitle

\section{Introduction}\label{sec:intro}

Algols are short orbital period (from several hours to tens of days)
semi-detached binaries composed of a hot B-A main sequence star, and a cool
F-K giant star \citep{1955AnAp...18..379K,1983ApJS...52...35G}. Because of
the small separation, the initially more massive star (henceforth the
donor) overfills its Roche lobe and matter then channels to the companion
(the gainer) through the inner Lagrangian point $\mathcal{L}_{1}$, i.e., by
Roche Lobe Over-Flow (RLOF). The duration of the mass-transfer phase (with
rates reaching $\approx 10^{-4}$~M$_{\odot}$~yr$^{-1}$) is long enough
(several $10^{5}$~yrs) that a significant amount of mass can be accreted by
the gainer (up to several solar masses). Observations clearly attest to the
reversal of the mass ratio during RLOF with a donor star (now a giant) less
massive than its main-sequence companion. The global evolution is well
established but many open questions still remain such as discrepancies
between canonical simulations and observations about the final orbital
periods and mass ratios \citep{2010A&A...510A..13V, 2011A&A...528A..16V}.
	
Previous numerical simulations of Algols \citep{2001ApJ...552..664N,
  2011A&A...528A..16V} have provided a broad understanding of short-period
binaries, focusing on some specific aspects such as non-conservative mass
transfer \citep{1981A&A...103..111G} or the properties of the mass-transfer
stream \citep{1975MNRAS.170..325F}. However, except for the study of
\cite{2010MNRAS.406.1071D}, there are basically no simulations that follow
in detail the angular-momentum evolution for both the stars and the orbit,
and that account for torques arising from tides, mass transfer,
magnetic-field effects, disc accretion or direct impact and spin-down
mechanisms.

Moreover, the evolution of critically-rotating accretors is poorly
understood: it is well known that the transferred mass carries enough
angular momentum \citep{1967AcA....17..297K} to rapidly spin the gainer up
to its critical spin-angular velocity \citep{1981A&A...102...17P}
\begin{equation}
  \Omega_{\mathrm{crit,*}}=( G M_{*} /
  R_{*}^{3})^{1/2}. \label{eq:omegacrit}
\end{equation}
For a typical 6~+~3.6~M$_{\odot}$ system, with initial period
P$_{\mathrm{init}}$~=~2.5~days, in the absence of spin-down mechanisms,
only 3 per cent (0.12~M$_{\odot}$) of the total amount of matter
transferred by RLOF (more than 5~M$_{\odot}$) is enough to spin the gainer
up to the critical rotation. However, it is not clear whether the gainer
reaches the critical velocity, is spun down due to any braking mechanisms,
or if accretion stops \citep{2007AIPC..948..321D}. Many physical processes
have been proposed to spin down the gainer, including tides
\citep{1977A&A....57..383Z}, magnetic braking \citep{1996MNRAS.280..458A,
  2000A&A...353..227S, 2010MNRAS.406.1071D}, or to limit the accretion of
angular momentum through the interaction with an accretion disc
\citep{1991ApJ...370..604P, 1991ApJ...370..597P, 1991MNRAS.253...55C,
  1993A&A...274..796B} but none of them have been applied to a binary-star
evolution code that consistently follows the stellar spin velocity, as well
as all torques owing to mass accretion and mass loss.

It has also been shown that some Algol systems must be non-conservative to
comply with observations of mass ratios (for example
\citealt{1974A&A....36..113R, 1975MmSAI..46..217M, 1980A&A....83..217M,
  1993MNRAS.262..534S, 2001ApJ...552..664N}). Yet, only mass loss through
the outer Lagrangian point $\mathcal{L}_{3}$ \citep{2007ARep...51..836S},
bipolar jets \citep{2007A&A...463..233A} or hotspot formation
\citep{2011A&A...528A..16V} have been proposed to account for this kind of
evolution. Non conservative mass transfer in stellar evolution codes is
generally treated via a free parameter defining the amount of matter
accreted by the gainer star compared to the amount of matter passing
through the inner Lagrangian point \citep{1993A&AS...97..527D,
  1998MNRAS.297..760S, 2002ApJ...575..461E, 2010MNRAS.406.1071D}.

In this article, we focus on the evolution of the gainer's spin-angular
velocity and non-conservative evolution. To do so, we study a specific case
B mass-transfer system (M$_{1}$~=~3.6~M$_{\odot}$, M$_{2}$~=~6~M$_{\odot}$,
and P$_{\mathrm{orb}}$~=~2.5~days) in which the donor is a hydrogen-shell
burning star. Mass transfer starts just after the star leaves the main
sequence and the rate is characterised by a short phase of rapid mass
transfer (hereafter rapid phase) where most of the mass is transferred,
followed by a long-lasting slow phase (hereafter quiescent phase) where
most of the orbital-separation changes take place.

To address the aforementioned problems (i.e. critical rotation of the
gainer and non-conservative evolution), we used the binary-star evolution
code \textsc{Binstar} \citep{2013A&A...550A.100S}.
Section~\ref{sec:binstar} presents the implemented binary input physics;
our treatment of rotation and the various spin-down mechanisms, as well as
the torques due to magnetic fields, star-disc interactions and tides. We
also describe our implementation of the hotspot physics and its
consequences on the mass accretion rate. In Sect.~\ref{sec:results}, we
apply these prescriptions to our specific Algol system and analyse the
effect of each torque on the gainer's spin. A calculation with a hotspot
configuration is also presented. In Sect.~\ref{sec:discussions},
observational constraints for Algol systems and related objects are
discussed, and their impact on the spin-down mechanism
investigated. Finally, we discuss the effect of non-conservative evolution
on the close environment of the system (i.e., the region impacted by the
system mass-loss), and examine the limitations of our assumptions regarding
solid-body rotation and disc formation. The conclusions are given in
Sect.~\ref{sec:conclusion}.

\section{The \textsc{Binstar} code}\label{sec:binstar}

\textsc{Binstar} is an extension of the 1-dimensional stellar evolution
code \textsc{Starevol} \citep{2006A&A...448..717S}, that handles the
simultaneous calculation of the binary orbital parameters (separation and
eccentricity) as well as the two stellar components. A complete description
of the code can be found in \cite{2013A&A...550A.100S}.

In this section, we outline our treatment of stellar rotation, and the
effects of mass transfer on the stellar spins. We specially focus on the
torque due to magnetic fields and the formation of a star-disc boundary
layer. Hereafter, the subscripts `g' and `d' refer to the gainer (primary
star in Algol systems), and donor (secondary star), respectively. The
models presented hereafter assume solid-body rotation. The limitations
associated with this assumption are discussed in Sect.~\ref{sec:rot}.

The RLOF mass-transfer rate follows the prescriptions described by
\cite{1988A&A...202...93R} and \cite{1990A&A...236..385K}. In their
formalism, two regimes have been considered, depending on whether the Roche
radius lies in the optically thick or thin layers of the donor star.

Let us finally emphasize that the (solid) rotation profile and all the
torques are computed at each iteration during the convergence process,
along with the evolution of the separation and mass transfer rates. This
procedure brings numerical stability and ensures that the calculation of
the orbital elements are in step with the stellar structures.

\subsection{Treatment of angular momentum in \textsc{Binstar}}

\textsc{Binstar} follows the angular-momentum evolution of each star and
the orbit. The total angular momentum of the system ($J_{\Sigma}$) is given
by
\begin{equation}
  J_{\Sigma} = J_{\mathrm{d}} + J_{\mathrm{g}} + J_{\mathrm{orb}} ,
\end{equation}
where $J_{\mathrm{d,g}}$ are the stellar spin-angular momenta and
$J_{\mathrm{orb}}$ is the orbital component
\begin{equation}
  J_{\mathrm{orb}}=M_{\mathrm{d}}M_{\mathrm{g}}\left[\frac{Ga(1-e^{2})}{M_{\Sigma}}
  \right]^{1/2} , \label{eq:jorb}
\end{equation}
where $a$ is the binary separation, $e$ is the eccentricity and
$M_{\Sigma}=M_{\mathrm{d}}+M_{\mathrm{g}}$ is the total mass of the
system. Evolution is conservative if $M_{\Sigma}$ and $J_{\Sigma}$ are constant
during the evolution. The conservation of angular momentum implies that
\begin{equation}
  \dot{J}_{\mathrm{orb}}=\dot{J}_{\Sigma}-\,\dot{J}_{\mathrm{d}}-\,
  \dot{J}_{\mathrm{g}} , \label{eq:cons_jorb}
\end{equation}
where $\dot{J}_{\mathrm{d,g}}$ are the torques applied on each star,
$\dot{J}_{\mathrm{orb}}$ is the rate of change of the orbital angular
momentum, and $\dot{J}_{\Sigma}$ represents the angular-momentum loss rate
by the system in non-conservative evolution. Torques from mass transfer and
tides are detailed in \cite{2013A&A...550A.100S} and summarised in
Table~\ref{tab:tab_ang_budget}. 

\begin{table*}
  \caption{Summary of the angular-momentum transfer rates involved in
    a non-conservative system ($\beta<1$). The parameter
    $j_{\mathrm{impact}}$ ($j_{\mathrm{disc}}$) is the
    specific angular momentum computed with ballistic motion equations
    (boundary-layer treatment) and all $\dot{M}$ are defined positive. 
    The super-scripts `Wind' (`RLOF') refer to 
    stellar wind (Roche Lobe OverFlow) contributions while the
    subscripts `acc' (`loss') denote mass accretion (loss). Torques
    with the super-script `\textbf{B}' denote torques due to the
    presence of a magnetic field (see text for details). Also, $I$ is the
    star's momentum of inertia, $f_{\mathrm{jacc}}$ is a free parameter
    controlling the specific angular momentum accreted by wind accretion
    \citep{1944MNRAS.104..273B} and $f_{\Sigma}$ is another free
    parameter (set to 1 for the simulations presented in this paper)
    setting the specific angular momentum carried by the
    matter leaving the system, in units of the specific angular
    momentum of the binary system $j_{\mathrm{orb}}$.}
  \label{tab:tab_ang_budget}
  \begin{center}
    \begin{tabular}{|@{}c@{}|@{}c@{}|@{}c@{}|@{}c@{}|}

      \cline{2-4}
      \multicolumn{1}{c|}{}&&&\\
      \multicolumn{1}{c|}{}& DONOR & GAINER & SYSTEM \\
      \multicolumn{1}{c|}{}&&&\\
      \hline
      &&&\\
      \begin{tabular}{c}
        $\dot{J}$\\
        LOSS \\
      \end{tabular}
      & 
      \begin{tabular}{@{}c@{}c@{}c@{}}
        $\dot{J}_{\mathrm{loss,d}}^{\mathrm{wind}}$&=&$-\frac{2}{3}
        \dot{M}_{\mathrm{loss,d}}^{\mathrm{wind}}
        \Omega_{\mathrm{spin,d}}
        R_{\mathrm{d}}^{2}$\\ [1em]

        $\dot{J}_{\mathrm{loss,d}}^{\mathrm{RLOF}}$&=&$-
        \dot{M}_{\mathrm{loss,d}}^{\mathrm{RLOF}}
        \Omega_{\mathrm{spin,d}}
        R_{\mathrm{d}}^{2}$\\ [1em]

        $\dot{J}_{\mathrm{d}}^{\mathrm{tides}}
        $&=&$ - I_{\mathrm{d}}
        \dot{\Omega}_{\mathrm{spin,d}}^{\mathrm{tides}}$ \\
	
      \end{tabular}	
      & 
      \begin{tabular}{@{}c@{}c@{}c@{}}
        $ \dot{J}_{\mathrm{loss,g}}^{\mathrm{wind}}$& = &$-
        \frac{2}{3} \dot{M}_{\mathrm{loss,g}}^{\mathrm{wind}}
        \Omega_{\mathrm{spin,g}} R_{\mathrm{g}}^{2}
        ~~\textrm{if\ no\ }\vec{B}$~~~~\\ [1em]
        
        $\dot{J}_{\mathrm{loss,g}}^{\mathrm{disc,}\vec{B}}$& = &$
        \frac{-\mu_{B}^{2} \Omega_{\mathrm{spin,g}}^{2} }{ 3 G
          M_{\mathrm{g}}}$ \\[1em]
        
        ~~~~$\dot{J}_{\mathrm{loss,g}}^{\mathrm{wind,}\vec{B}}$&=&
        $\left\{
        \begin{array}{l}
          -\left[\left.\dot{M}_{\mathrm{loss,g}}^{\mathrm{wind}}\right.^{(4m-9)}
            \times \right.\\ \left. \frac{B_{*}^{8} R^{8m}}{(2 G
              M)^{2}} \right]^{\frac{1}{(4m-5)}}
          \Omega_{\mathrm{spin,g}}\\
        \end{array}
        \right.$\\ [1em]

        $\dot{J}_{\mathrm{g}}^{\mathrm{tides}} $&=& -
        $I_{\mathrm{g}}
        \dot{\Omega}_{\mathrm{spin,g}}^{\mathrm{tides}}$ \\
      \end{tabular}
      & 
      \begin{tabular}{@{}c@{}c@{}c@{}}
       ~~~~ $\dot{J}_{\mathrm{orb,d}}^{\mathrm{wind}} $&=&$ - f_{\Sigma}
        (\dot{M}^{\mathrm{wind}}_{\mathrm{loss,d}} -
        \dot{M}^{\mathrm{wind}}_{\mathrm{acc,g}}) j_{\mathrm{orb}}$~~~~\\[1em]

        ~~~~$\dot{J}_{\mathrm{orb,g}}^{\mathrm{wind}}$& = &$ - f_{\Sigma}
        (\dot{M}^{\mathrm{wind}}_{\mathrm{loss,g}} -
        \dot{M}^{\mathrm{wind}}_{\mathrm{acc,d}})
        j_{\mathrm{orb}}$~~~~\\[1em]
        
        ~~~~$\dot{J}_{\mathrm{orb}}^{\mathrm{RLOF}} $&
        = &$ - f_{\Sigma} (1-\beta)
        \dot{M}^{\mathrm{RLOF}}_{\mathrm{loss,d}} j_{\mathrm{orb}}$~~~~\\
      \end{tabular}
      \\
      &&&\\
      \hline
      &&&\\
      
      \begin{tabular}{c}
        $\dot{J}$\\
        GAIN \\
      \end{tabular}
      &
      \begin{tabular}{@{}c@{}c@{}c@{}}
        ~~~~$\dot{J}_{\mathrm{acc,d}}^{\mathrm{wind}} $&=&$ +\frac{2}{3}
        f_{j_{\mathrm{acc}}}
        \dot{M}_{\mathrm{acc,d}}^{\mathrm{Bondi-Hoyle}}
        \Omega_{\mathrm{spin,g}} R_{\mathrm{g}}^{2}$ ~~~~\\
      \end{tabular}
      &
      \begin{tabular}{@{}c@{}c@{}c@{}}
        $\dot{J}_{\mathrm{acc,g}}^{\mathrm{wind}} $&=&$ +\frac{2}{3}
        f_{j_{\mathrm{acc}}}
        \dot{M}_{\mathrm{acc,g}}^{\mathrm{Bondi-Hoyle}}
        \Omega_{\mathrm{spin,d}} R_{\mathrm{d}}^{2}$\\ [1em]
        
        $\dot{J}_{\mathrm{acc,g}}^{\mathrm{RLOF}} $&=&$\left\{
        \begin{array}{l} 
          \mathrm{for\ direct\ impact:}\\ \beta
          \dot{M}_{\mathrm{loss,d}}^{\mathrm{RLOF}}
          j_{\mathrm{impact}}\\ \mathrm{
            for\ disc\ accretion:}\\ \beta
          \dot{M}_{\mathrm{loss,d}}^{\mathrm{RLOF}}
          j_{\mathrm{disc}}\\
        \end{array}
        \right.$\\
        
      \end{tabular}
      &
      \begin{tabular}{@{}c@{}}
        No gain for the system \\
      \end{tabular}
      \\
      &&&\\
      \hline
    \end{tabular}
  \end{center}
\end{table*}

\subsection{Magnetic torques}

Depending on its configuration, a magnetic field can either spin the star
up or down. We consider two effects: magnetic-wind braking and
disc-locking.

\subsubsection{Magnetic-wind braking}\label{sec:mag_wb}

Stars expel matter through winds. Assuming the specific angular
  momentum of the surface layers of the star to be
\begin{equation}
  j_{\mathrm{surf}} = \Omega_{*} R_{*}^{2},
\end{equation}
this matter removes angular momentum at a rate
\begin{equation}
  \dot{J}_{\mathrm{loss,*}}^{\mathrm{wind,}\vec{B}}
  = -\vert\dot{M}_{\mathrm{W}} \vert\Omega_{*}
  R_{*}^{2} < 0,
\end{equation}
where $\dot{M}_{\mathrm{W}}$ is the wind mass-loss rate, $R_{*}$ is the
stellar radius and $\Omega_{*}$ is its spin-angular speed. If the magnetic
field co-rotates with the star, particles emitted by the wind can either be
trapped within the dead zone (where the magnetic-field lines are closed),
or leave the system by flowing along open magnetic-field lines up to the
Alfv\`{e}n radius $R_{\mathrm{A}}$
\citep{1967ApJ...148..217W,1968MNRAS.138..359M}. Beyond $R_{\mathrm{A}}$
where the wind speed equals the Alfv\`{e}n speed, matter decouples from
the magnetic field and particles can freely escape the system. The
angular-momentum loss rate at $R_{\mathrm{A}}$ is given by
\begin{equation} \label{eq:AMrate}
  \dot{J}_{\mathrm{loss,*}}^{\mathrm{wind,}\vec{B}} = -
  \vert\dot{M}_{\mathrm{W}}\vert\Omega_{*} R_{\mathrm{A}}^{2}.
\end{equation}
The torque is efficient because $R_{\mathrm{A}} \gg R_{*}$ and can be written
as (e.g. \citealt{2010MNRAS.406.1071D})
\begin{equation}
  \dot{J}_{\mathrm{loss,*}}^{\mathrm{wind,}\vec{B}}
  =- \left[\vert \dot{M}_{\mathrm{W}}\vert^{(4m-9)}B_{*}^{8}(2
    G M_{*})^{-2} R_{*}^{8m}\right]^{1/(4m-5)}\Omega_{*}.
  \label{eq:mdot_wind}
\end{equation}
where $m$ is a parameter characterising the topology of the magnetic field
of magnitude $B_{*}$ at the stellar surface. For our purpose, we assume a
bipolar configuration ($m=3$) which is a reasonable approximation around
the Alfv\`{e}n radius where the matter is released
\citep{1992MNRAS.259P..23L}. In our models, only the magnetic-field
strength $B_{*}$ is left as a free parameter. The wind mass-loss rate is
determined using the prescription of \cite{1975psae.book..229R}. In a case
study, we also considered a rotationally enhanced wind mass loss where the
modified rate is given by \citep{2001A&A...373..555M}
\begin{equation}
  \frac{\dot{M}^{\mathrm{wind}}(\Omega)} {\dot{M}^{\mathrm{wind}}(0)} =
  \left( \frac{1-\Gamma}{1-\frac{\Omega^{2}}{2 \pi
        G \rho_{m}} - \Gamma} \right)^{\frac{1}{\alpha} - 1}, 
\label{eq:wind_rot}
\end{equation}
where $\Gamma$ is the ratio of the stellar luminosity over the Eddington
luminosity (Eq.~\ref{eq:Ledd}), $\rho_{m}$ the stellar mean density
\begin{equation}
  \rho_{m} = \frac{3 M_{*}}{4 \pi * R_{*}^{3}},
\end{equation} 
and $\alpha$ a force multiplier such that $\alpha$ = 0.52, 0.24, 0.17 and
0.15, for log(T$_{\mathrm{eff}}$ $\ge$ 4.35, 4.30, 4.00 and 3.90
respectively \citep{1995ApJ...455..269L}.

\subsubsection{Disc-locking}\label{sec:disc_lock}
			
If an accretion disc is present, the gainer's magnetic field anchors into
it. Since the disc does not co-rotate with the star, magnetic-field lines
are twisted, creating a toroidal magnetic-field component that generates a
torque on the star. If the magnetic field is anchored beyond the
co-rotation radius defined as
\begin{equation}\label{eq:rc}
  R_{\mathrm{co}}=\left( \frac{G M_{*}}{\Omega_{*}^{2}} \right)^{1/3} ,
\end{equation}
the torque is negative and the star is spun down. The inner disc is
truncated at the Alfv\'{e}n radius $R_{\mathrm{A}}$, because below this
region, the magnetic pressure is larger than the gas pressure.
			
We assume a simple power law for the magnetic-field strength of the form
$B(r) = B_{*}(R_{*}/r)^{m}$ for $r > R_{\mathrm{A}}$,
considering the same magnetic-field geometry parameter as before ($m=3$ for
a bipolar configuration). The resulting torque is given by
\citep{1996MNRAS.280..458A}
\begin{equation}
  \dot{J}_{\mathrm{loss,*}}^{\mathrm{disc,}\vec{B}} =
  \frac{\mu^{2}}{3}(R_{\mathrm{A}}^{-3}-2 R_{\mathrm{co}}^{-3/2}
  R_{\mathrm{A}}^{-3/2}) , 
\label{eq:j_disc_locking}
\end{equation}
with $\mu=B_{*}R_{*}^{3}$. Depending on the location of $R_{\mathrm{co}}$
and $R_{\mathrm{A}}$, the global torque can be either positive or negative
and is most efficient at spinning-down the gainer when $R_{\mathrm{A}} =
R_{\mathrm{co}}$ because in this configuration there are no regions of the
disc that rotate faster than the star (remember that below $R_A$ the disc
is truncated). The expression of this maximum torque
\citep{2000A&A...353..227S} is given by
\begin{equation}\label{eq:disc_lock_torque_final}
  \dot{J}_{\mathrm{loss,*}}^{\mathrm{disc,}\vec{B}} =
  -\frac{\mu^{2} \Omega_{*}^{2}} {3 G M_{*}}.
\end{equation}
and does not depend on the disc properties such as its mass.

\subsection{Tidal effects}

Tides have two effects on a binary system: they circularise the orbit of
eccentric systems, and synchronise the stellar spin with the orbital
period. The implementation of tides in \textsc{Binstar} follows the
prescription described by \cite{1977A&A....57..383Z} and includes the
\cite{1989A&A...220..112Z} refinement for convective stars in short-period
systems (see \citealt{2013A&A...550A.100S} for details).

\subsection{Accretion disc and star-disc boundary layer}
\label{sec:bl}

The matter-stream escaping the donor through the $\mathcal{L}_{1}$ point will
free-fall into the gainer's potential well. If the stream's minimum
distance of approach to the gainer is such that
$R_{\mathrm{min}}>R_{\mathrm{g}}$, then the stream will orbit the gainer
until it collides with itself. Further collisions between particles lead to
the formation of a ring at the circularisation radius. The matter will
then spread out and an accretion disc forms
\citep{1967AcA....17..297K,1975ApJ...198..383L}.

The properties of accretion discs around fast-rotating stars have been
widely studied (e.g. \citealt{1991ApJ...370..597P, 1991ApJ...370..604P,
  1991MNRAS.253...55C,1993A&A...274..796B}). Their 1D calculations show that
accretion does not stop when the gainer reaches its critical velocity
because of the presence of a boundary layer between the star and the
disc. The physics of this boundary and its implementation in
\textsc{Binstar} are presented in the next section.

\subsubsection{The Paczynski model}

The main idea of this model \citep{1991ApJ...370..597P} is to assume that
the star and the disc behave as one fluid. A polytropic equation of state
is used for both the disc and the star with the pressure given by
\begin{equation}\label{eq:bl_poly}
  P = K \rho^{ 1 + 1/n}
\end{equation}
where $K$ and $n$ are two constants. The hydrostatic equilibrium of the
disc is ensured by balancing the gravitational acceleration, centrifugal
acceleration and pressure gradient (in a cylindrical coordinate system in
which $r$ is the distance from the rotation axis and $z$ the distance from
the equatorial plane), i.e.
\begin{equation}\label{eq:bl_hydrostatic1}
  r\Omega^{2} = \left.\frac{\partial \Psi}{\partial r}\right|_{z} + 
  \left.\frac{\partial \Psi}{\partial z}\right|_{r}
  \frac{\mathrm{d}h}{\mathrm{d}r},
\end{equation}
where $z = h(r)$ is the disc thickness as a function of $r$, $\Omega$ is
the disc angular velocity and
\begin{equation} 
  \Psi = -\frac{G M_{*}}{ (r^{2} + z^{2})^{1/2}}
\end{equation}
is the gravitational potential of a point mass, $M_{*}$ being the stellar
mass. For a thin disc [$h(r)\ll r$], the equation of hydrostatic
equilibrium in the $z$-direction is 
\begin{equation}
\frac{1}{\rho} \frac{\partial P}{\partial r} = \left.\frac{\partial
  \Psi}{\partial z}\right|_{r} \approx -\frac{GM_{*}}{r^{3}}h(r).
\end{equation}
After some algebra, and using Eq.~(\ref{eq:bl_hydrostatic1}), the condition
for hydrostatic equilibrium may be re-written as
\begin{equation}\label{eq:bl_hydrostatic}
  \frac{h}{r}\frac{\mathrm{d} h}{\mathrm{d} r} = \frac{(r^{2} +
    h^{2})^{3/2}\Omega^{2}}{ GM_{*}} - 1
\end{equation}
The second equation expressing the angular-momentum conservation writes
\begin{equation}\label{eq:bl_AM_balance_1}
  \dot{M}\frac{\mathrm{d}j}{\mathrm{d}r} =
  \frac{\mathrm{d}\Gamma}{\mathrm{d} r} ,
\end{equation}
where $\dot{M}$ is the mass flux in the disc, $j$ is the specific angular
momentum carried with the disc material and $\Gamma$ is the torque due to
viscous stresses inside the disc. Using the polytropic equation of state,
the $\alpha$-prescription for the disc viscosity
\citep{1973A&A....24..337S} and the equation of the torque acting between
two cylinders in a differentially rotating disc
\citep{1981ARA&A..19..137P}, Eq.~(\ref{eq:bl_AM_balance_1}) is rewritten as
\citep{1991ApJ...370..597P}
\begin{equation}\label{eq:bl_AM_balance}
  \left[ 4\pi \alpha \frac{(G M_{*})^{n+1/2}}{K^{n}}\right]\frac{\mathrm{d}
    \Omega}{\mathrm{d} r} = \frac{r^{3n-3/2}}{z^{2n+3}}(\dot{J} -
  \dot{M}r^{2} \Omega).
\end{equation}

\subsubsection{Dimensionless equations} 
For convenience, dimensionless variables are introduced
\begin{equation} 
x = \frac{r}{R_{*}} ~~,~~ y = \frac{h}{R_{*}},
\end{equation}
\begin{equation} 
  \omega^{2} = \Omega^{2}\frac{R_{*}^{3}}{G M_{*}} ~~,~~
  \omega_{*}^{2} = \Omega_{*}^{2}\frac{R_{*}^{3}}{G M_{*}} ~~,~~
  \omega_{\mathrm{crit}}^{2} = \Omega_{\mathrm{crit}}^{2}\frac{R_{*}^{3}}{G M_{*}},
\end{equation}
where $\Omega_{*}$ and $\Omega_{\mathrm{crit}}$ are the stellar and
critical angular velocities respectively. We also define the two constants
\begin{equation}
  \varsigma \equiv 4 \pi \alpha \frac{G^{2}M^{2}}{\dot{M_{*}}K^{1.5}} ,
\end{equation}
and
\begin{equation}
  j_{*} \equiv \frac{\dot{J}}{\dot{M}( G M_{*} R_{*})^{1/2}} ,
\end{equation}
where $j_{*}$ represents the specific angular momentum of the disc material
at the effective stellar radius $R_{*}$ defined as
\citep{1939isss.book.....C}
\begin{equation}
  R_{*} = \frac{K}{0.4242GM_{*}^{1/3}}.
\end{equation}
Following \cite{1991ApJ...370..597P}, we use $\varsigma=10^{6}$ throughout
the computations.  In dimensionless form, Eqs.~(\ref{eq:bl_hydrostatic})
and (\ref{eq:bl_AM_balance}) can be re-expressed as
\begin{eqnarray}\label{eq:disc1}
  \frac{y}{x}\frac{\mathrm{d}y}{\mathrm{d}x} & =
  &\omega^{2}(x^{2}+y^{2})^{3/2} - 1 ,
  \\\label{eq:disc2} \frac{\mathrm{d}\omega}{\mathrm{d}x} & = &
  \frac{x^{3}}{ \varsigma y^{6}}(j_{*}-\omega x^{2}) .
\end{eqnarray}
This system of equations does not depend on the disc properties such as
mass or radial extent. This system is solved using a two-points boundary
value solver \citep{Capper}. For numerical reasons, two sets of boundaries
are used depending on the stellar spin-angular velocity.

For rapidly rotating stars, with $\omega_{*} / \omega_{\mathrm{crit}}
\approx 1$, the structure of the disc must satisfy the boundary
condition
\begin{equation}
  \omega = \omega_{*}~\mathrm{at}~r = r_{\mathrm{tr}} ,
\end{equation}								
where $r_{\mathrm{tr}} = 0.8 r_{\mathrm{polar}}$ is an arbitrary transition
radius between the star and the disc, and $r_{\mathrm{polar}}$ is the
star's polar radius.

For slowly rotating stars, the spin-angular velocity profile presents a
local maximum in the boundary layer. This maximum, located at a radius
$r_{\mathrm{nt}}$, corresponds to a zero-torque condition between the star
and the disc and the inner boundary condition writes
\begin{equation}				
  \left( \frac{\mathrm{d}\omega}{\mathrm{d}r} \right)_{r_{\mathrm{nt}}} = 0 .
\end{equation}
A Runge-Kutta integrator \citep{Press} is used to solve the inner part of
the system, between $r_{\mathrm{nt}}$ and $r_{\mathrm{tr}}$.

For both rapidly and slowly rotating systems, the outer boundary condition
is
\begin{equation}
  \frac{\mathrm{d}\omega}{\mathrm{d}r} =
  \frac{\mathrm{d}\omega_{\mathrm{Kepler}}}{\mathrm{d}r} ,
\end{equation}
where $\omega_{\mathrm{Kepler}}$ is the Keplerian angular speed at the
outer edge of the disc. Details on the method of calculation can be found
in \cite{1991ApJ...370..597P}.

The solution of this system of equations provides the value of the specific
angular momentum $j_{\mathrm{disc}} = j_{*}$ accreted by the star as a
function of its stellar spin. The torque exerted on the star is then
\begin{equation}
  \dot{J}_{\mathrm{acc,g}}^{\mathrm{RLOF}} =
  \vert \dot{M}_{\mathrm{acc,g}}^{\mathrm{RLOF}} \vert j_{\mathrm{disc}},
\end{equation}
where $\dot{M}_{\mathrm{acc,g}}^{\mathrm{RLOF}}$ is the mass accretion rate
for the gainer star. Note that $j_{\mathrm{disc}}$ can be negative.

To ease the computation when the star is not critically rotating (less than
80\% of the critical spin), the accreted specific angular momentum can be
set as a fraction $f_{\mathrm{j_{\mathrm{disc}}}}$ of the Keplerian
specific angular momentum at the surface of the star, i.e.
\begin{equation}
  j_{\mathrm{disc}} = f_{\mathrm{j_{\mathrm{disc}}}} \sqrt{G M_{\mathrm{g}}
    R_{\mathrm{g}}} .
\end{equation}
We use $f_{\mathrm{j_{\mathrm{disc}}}} = 1$, which is the mean value
computed with the `boundary-layer' mechanism for non-critically rotating
accretors.

In the disc, advection of matter occurs due to the transport of angular
momentum by viscosity. By treating the star-disc boundary layer as one
fluid, the same mechanism also applies at the stellar surface: when the
star reaches critical rotation, the accretion of angular momentum is no
longer possible and viscous processes remove spin-angular momentum
from the star, keeping it at a critical value but not exceeding it. This
mechanism allows the star to accrete large amounts of mass while giving
angular momentum back to the disc. This extra angular momentum leads to the
spreading of the disc. The outer disc radius is truncated because of tidal
interaction between the disc and the donor star
\citep{1979MNRAS.186..799L,1988MNRAS.232...35W,1994PASJ...46..621I}, and
angular momentum is returned to the orbit.

\subsection{Direct impact and hotspot formation}
\label{sec:MOD:hot}

A second mode of mass transfer occurs when
$R_{\mathrm{min}}<R_{\mathrm{g}}$, such that the stream impacts the
gainer's surface (hereafter `direct impact'). In this case, we follow the
evolution of the stream through a ballistic approach to determine the
precise location, velocity and angular momentum of the impacting stream
\citep{1963ApJ...138..481H,1964AcA....14..231K,1975MNRAS.170..325F}.

When the particle collides with the star, its specific angular momentum is
\begin{equation}
  j_{\mathrm{impact}}= \Vert \vec{r} \wedge \vec{v} \Vert
  , \label{eq:jacc}
\end{equation}
where $\vec{r}$ and $\vec{v}$ are respectively the radius vector of the
particle from the gainer's stellar centre, and the particle velocity at the
impact location. These are determined by integrating the equations of
motion of the particle using a Runge-Kutta scheme. Using Eq.~(\ref{eq:jacc}),
the torque applied on the accretor is given by
\begin{equation}
  \dot{J}_{\mathrm{acc,g}}^{\mathrm{RLOF}}= \vert
  \dot{M}_{\mathrm{acc,g}}^{\mathrm{RLOF}} \vert j_{\mathrm{impact}} .
\label{eq:torque_RLOF_acc_impact}
\end{equation}
When the stream impacts the star (or the disc's outer radius), its kinetic
energy is converted into thermal energy and a hotspot forms
\citep{2004AN....325..225P,2012ApJ...750...59L}. If the hotspot luminosity
exceeds the Eddington value
\begin{equation}
  \label{eq:Ledd}
  L_{\mathrm{edd}} = \frac{4 \pi c G M_{*}}{\kappa},
\end{equation}
where $\kappa$ is the plasma opacity, the matter escapes the system. We
show in Appendix~\ref{ap:hotspot} (all quantities entering
Eq.~(\ref{eq:Macc_crit}) being defined there) that if the mass-transfer rate
exceeds
\begin{equation}
  \dot{M}_{\mathrm{acc}}^{\mathrm{crit}}={ \frac{2}{v_{\mathrm{imp}}^{2} +
      a_{\mathrm{g}}^{2} \Omega_{\mathrm{orb}}^{2}}}\left(
    \frac{\tilde{L}_{\mathrm{edd,g}}-L_{\mathrm{g}}}{\tilde{K}}
    + \frac{E_{\mathrm{rot}}^{\prime}-E_{\mathrm{rot}}}{\Delta
      t}\right) ,
\label{eq:Macc_crit}
\end{equation}
the evolution becomes non-conservative
\citep{2000NewAR..44..111E,2008A&A...487.1129V}. We use a prescription for
$\tilde{K}$ depending only on the total mass of the system and based on
observations (see Appendix~\ref{ap:hotspot}, Eq.~(\ref{eq:ktilde})).

The parameter $\beta$ defining the fraction of the transferred mass that
escapes the system is given by
\begin{equation}
  \beta = \left| \frac{\dot{M}_{\mathrm{g,acc}}^{\mathrm{RLOF}}}{
\dot{M}_{\mathrm{d,loss}}^{\mathrm{RLOF}}}\right| ,
\end{equation}
where $\dot{M}_{\mathrm{g,acc}}^{\mathrm{RLOF}} =
\min(\dot{M}_{\mathrm{loss,d}}^{\mathrm{RLOF}},
\dot{M}_{\mathrm{acc}}^{\mathrm{crit}})$. Note that the wind mass-loss
contribution is not included.

In short-period binaries, matter may also escape the system through the
third Lagrangian point \citep{2007ARep...51..836S} at a rate of
approximately $10^{-10}$ M$_{\odot}$ yr$^{-1}$, which is higher than
stellar winds but still negligible compared to the RLOF mass-transfer
rate. We neglect this aspect in our present investigation.

\section{Results}\label{sec:results}
	
In this section, we present new calculations of Algol evolution, taking the
previously described spin-down mechanisms and non-conservative evolution
into account. Our reference system is a 6~M$_{\odot}$ + 3.6~M$_{\odot}$
with initial period of 2.5 days. Both stars are initially relaxed on the
zero-age main-sequence with composition $X_{\mathrm{g}}$ = 0.735 and
$X_{\mathrm{d}}$ = 0.695 and are synchronised with the orbit. Initial
effective temperatures (luminosities) are $\approx$10500~K
($\approx$130~L$_{\odot}$) for the gainer and $\approx$19500~K
($\approx$1030~L$_{\odot}$) for the donor. In this study, we do not
consider Bondi-Hoyle accretion
($\dot{J}_{\mathrm{acc,[g,d]}}^{\mathrm{wind}}$, see
Table~\ref{tab:tab_ang_budget}) because of its negligible impact compared
to Roche lobe overflow mass transfer. By default, no spin-down mechanisms
are used unless mentioned.

\subsection{On the spin up of the gainer}\label{sec:spin_up}
	
\begin{figure}[t!]
  \centering \includegraphics[width=.48\textwidth]{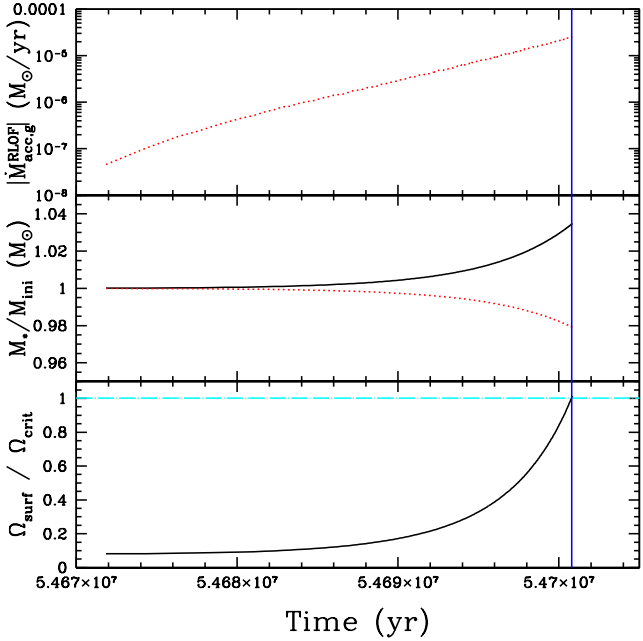}
  \caption{Evolution of the mass-transfer rate (top), stellar masses
    (middle), surface velocity (bottom) for a 6~M$_{\odot}$ donor (Solid
    black line) and 3.6~M$_{\odot}$ gainer (dotted red line) system with
    initial period P$_{\mathrm{init}}$~=~2.5~days.}
  \label{fig:mdot_onset}
\end{figure}
		
Figure~\ref{fig:mdot_onset} shows the evolution of the mass-transfer rate
(top panel), stellar masses (middle panel) and the surface-spin velocity of
the gainer (in units of the critical Keplerian velocity; bottom panel) for
our reference system, undergoing case B mass transfer. Here, we have not
considered any spin-down mechanisms and have assumed fully conservative
mass transfer. Mass and angular-momentum losses from the system only occur
via winds (quasi-conservative evolution).

During the rapid phase, the mass-accretion rate reaches values up to
$10^{-4}$~M$_{\odot}$~yr$^{-1}$, and by about 3$\times$10$^{4}$~years, the
gainer reaches critical velocity (blue vertical line in
Fig.~\ref{fig:mdot_onset}). At this point we stop the evolution. During
this period, only a small amount of matter is transferred, the gainer's
mass increases by 0.12~M$_{\odot}$ which represents 3\% of its initial mass
(Fig.~\ref{fig:mdot_onset}, middle panel). However, if the evolution was
continued, more than 5~M$_{\odot}$ would have been transferred.

\subsection{Effect of magnetic fields}\label{sec:RES:mag}

\begin{figure}[t!]
  \centering
  \includegraphics[width=.48\textwidth]{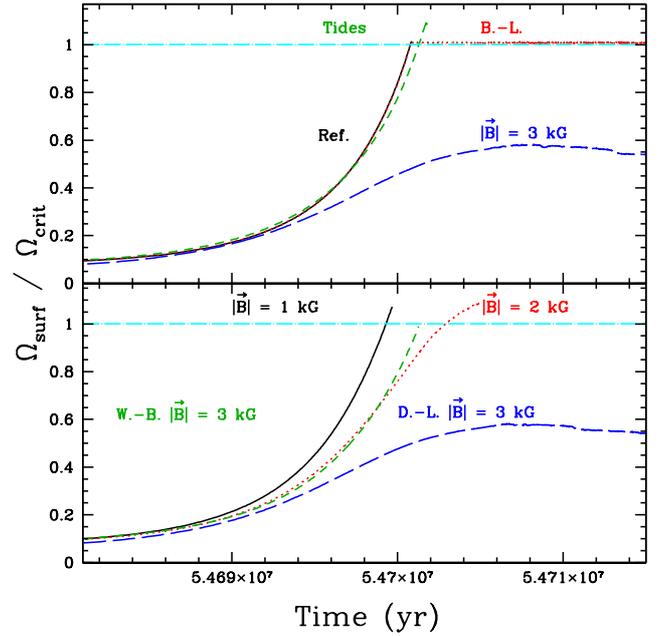}
  \caption{Evolution of the gainer's surface velocity (normalised to the
    critical velocity) for different spin-down mechanisms for our 6 +
    3.6~M$_{\odot}$ system with initial period P$_{\mathrm{init}}$ = 2.5
    days. The horizontal line represents the critical spin-angular rotation
    of the star. All simulations are stopped when $\Omega_{\mathrm{surf}}
    \approx \Omega_{\mathrm{crit}}$. {\bf Top panel:} solid black line
    (labelled `Ref.'): $|\vec{B_{*}}|$~=~0~kG, no braking mechanism; dotted
    red line (labelled `B.-L.'): boundary-layer mechanism, no magnetic
    field; dashed green line: \cite{1989A&A...220..112Z} tidal effect
    treatment only; long-dashed blue line: magnetic braking with
    $|\vec{B_{*}}|$~=~3~kG. {\bf Bottom panel:} solid black line: magnetic
    braking with $|\vec{B_{*}}|$~=~1~kG; dotted red line: magnetic braking
    with $|\vec{B_{*}}|$~=~2~kG; dashed green line (labelled M.-W.):
    magnetic-wind braking only with $|\vec{B_{*}}|$~=~3~kG; long-dashed
    blue line (labelled D.-L.): disc locking only with
    $|\vec{B_{*}}|$~=~3~kG.}
  \label{fig:B_vsurf_vcrit}
\end{figure}

The efficiency of magnetic-wind braking strongly depends on the mass-loss
rate of the gainer (and, to a certain extent, on the parameter $\beta$ for
non-conservative evolution), and on $\vert \vec{B_{*}} \vert$. In the
following simulations, we assume that the disc-locking activates as soon as
the gainer is spun up and is at work even if direct impact is
expected. Although contradictory, this assumption does not conflict with
observation of binaries in which the gainer is surrounded by a disc even
though the \cite{1975ApJ...198..383L} criterion\footnote{A criterion
  defining whether a system is in the direct impact or disc accretion
  regime, based on the binary parameters (mass ratio and orbital
  separation).} implies direct impact (see Sect.~\ref{sec:classI}).

The top panel of Fig~\ref{fig:B_vsurf_vcrit} shows the evolution of the
gainer's surface spin-angular velocity (normalised to the critical value)
for different values of $\vert \vec{B_{*}} \vert$, where we consider both
magnetic-wind braking and disc-locking. Only a magnetic field of intensity
$\vert \vec{B_{*}} \vert \ga$~3~kG can spin down the star sufficiently to
prevent critical rotation. 

In the lower panel of Fig.~\ref{fig:B_vsurf_vcrit}, we present the
evolution of the normalised spin-angular velocity for each magnetic-braking
mechanism individually. As shown, wind braking is very inefficient but
various effects can contribute to strengthen its action. First, rotation
can substantially enhance the mass loss rate because of the reduced
gravity. To quantify this effect, we ran additional simulations using
Eq.~(\ref{eq:wind_rot}) for the wind loss rate.  Despite the fact that the
mass-loss rate is increased by a factor of almost 3, the 2~kG configuration
still produces a critically rotating gainer and in the 3~kG case, the
difference in stellar spin is insignificant, of the order of 2
percents. Therefore, because disc-locking is so efficient (and independent
of mass loss rate), change in the wind rate due to rotation has a
negligible impact.  Alternatively, \cite{1992MNRAS.256..269T} argue that
the dynamo process at the origin of the magnetic-field generation can
enhance the mass loss rate and hence the magnetic torque. Besides, the
evolution may not be conservative and mass may leave the system from the
hotspot (see Sects.~\ref{sec:hotspot_res} and \ref{sec:results:betaB}) or
from the outer Lagrangian points $\mathcal{L}_{2}$ and $\mathcal{L}_{3}$ if
the Roche lobe geometry is modified by radiation pressure
\citep{2009A&A...507..891D} for example. On the other hand, the
disc-locking torque has been overestimated because we set $R_{\mathrm{co}}
= R_{\mathrm{A}}$ in Eq.~(\ref{eq:j_disc_locking}) and considered that the
whole outer disc is contributing to spin down the star. This is not
necessarily the case because the disc magnetosphere can screen the stellar
magnetic field, preventing it to anchor into the disc and therefore to spin
the star down \citep{1979ApJ...232..259G}. Note also that not all Algol
systems possess a disc during the rapid mass-transfer phase, and therefore
the disc-locking mechanism may not operate universally.

Observations suggest that magnetic fields of 3~kG for main sequence stars
are rare even for Ap and Bp stars, which are known to be more magnetically
active than standard A-B stars typically found in Algols
\citep{2006AN....327..289H, 2007AN....328..475H, 2003A&A...407..631B,
  2009MNRAS.394.1338B}. However, these observations relate to slow
rotators, and we may expect the surface magnetic-field strength to have
been much larger when the star was spinning faster due to dynamo generation
during the rapid phase \citep{1999A&A...349..189S}. On the other hand, the
(initially) 3.6~M$_{\odot}$ gainer star has an extended radiative envelope
which limits the dynamo processes \citep{1983Ap&SS..90..217P}, so dynamo
effects might not be efficient. In addition, it is difficult to
observationally determine the magnetic-field strength of the main-sequence
gainer in Algols because it is likely weaker than that of the red-giant
gainer which possesses an expended convective envelope
\citep{2005ApJ...621..417R}. Given the large uncertainties surrounding the
modelling of these braking mechanisms, it is difficult to state whether or
not, a 3~kG magnetic field is a plausible value. Additional magnetic field
determinations are clearly needed to better understand the process(es) at
work in Algols.

\subsection{Effects of tides}
	
Tidal forces steeply increase with decreasing orbital separation and
therefore may be efficient in short-period Algols. The dashed green curve
of Fig.~\ref{fig:B_vsurf_vcrit} shows the evolution of the surface velocity
(in units of the critical velocity) for our 6 + 3.6~M$_{\odot}$ system,
including tidal braking. We confirm the results of
\cite{2010MNRAS.406.1071D} that tides are not efficient enough to spin-down
the gainer and compensate for the spin-up produced by angular-momentum
accretion because of the much longer tidal time-scale
($\tau_{\mathrm{sync}} \approx 10^{8}$~yr) compared to the mass-transfer
time-scale ($\tau_{\mathrm{acc}} = M_{\mathrm{g}} /
\dot{M}_{\mathrm{acc,g}}^{\mathrm{RLOF}} \approx 10^{4}$--$10^{5}$~yr).

Observations of Algols during the slow accretion phase indicate that they
are spinning slower than their critical velocity. In fact, most Algols are
synchronised with the orbital period \citep{1989SSRv...50..191W,
  1996MNRAS.283..613M}. Thus tides may help to spin-down the gainer during
the long-lasting quiescent phase, but they are too weak to maintain the
gainer below the critical spin velocity during the entire mass-transfer
episode.

\subsection{Star-disc boundary-layer treatment}

\begin{figure}[t!]
  \centering
  \includegraphics[width=.48\textwidth]{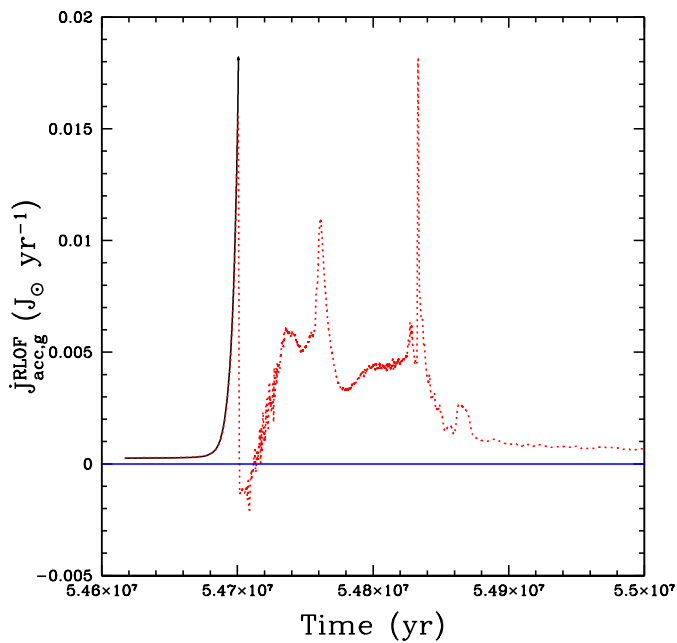}
  \caption{Evolution of the torque on the gainer star
    $\dot{J}_{\mathrm{acc,g}}^{\mathrm{RLOF}}$. Solid black line: No spin
    down mechanism (the simulation is stopped when the gainer reaches the
    critical rotation). Dashed red line: boundary-layer model. The
    horizontal line represents the limit where the net torque spins the
    star up (above) or spins it down (below).}
  \label{fig:jevol}
\end{figure}

Although disc formation in semi-detached binaries is determined by the
stellar and orbital properties of the system \citep{1975ApJ...198..383L},
we assume the formation of an accretion disc whenever the star reaches its
critical spin rate. Our paradigm is that the stellar matter which cannot be
accreted by the fast rotating and distorted star will surround the gainer,
eventually forming a disc (see Sect.~\ref{sec:classI}). In practice, the
treatment of the boundary layer activates once the star has reached 80\% of
its critical spin-angular velocity. As a result, the gainer is initially
spun up in the same way as if no braking mechanism was present and is then
kept at the critical spin-angular velocity when the boundary layer is at
work.

Figure~\ref{fig:jevol} shows the evolution of the torque applied on the
star due to mass accretion. When the boundary-layer mechanism is activated,
the torque quickly drops to zero when the star reaches the critical spin
velocity. Since the momentum of inertia of the donor keeps rising due to
the increase in mass and radius, angular momentum needs to be evacuated in
order to maintain the rotation at the critical value, hence the temporarily
negative value of $\dot{J}_{\mathrm{acc,g}}^{\mathrm{RLOF}}$
(Fig.~\ref{fig:B_vsurf_vcrit}). All the transferred matter is accreted and
the evolution of the system remains conservative.

When a disc forms because the separation is large enough, i.e. when the
gainer has a small filling-factor, the boundary model is activated and as
long as the gainer's surface velocity remains below
$0.9\Omega_\mathrm{crit}$, the accreted specific angular momentum is
basically equal to the Keplerian value
\begin{equation}
  j_{\mathrm{crit}} = \sqrt{G M_{*} R_{*,\mathrm{eq}}} ,
\end{equation}
where $R_{*,\mathrm{eq}}$ is the stellar equatorial radius. Note that our
computed values for $j_{\mathrm{crit}}$ are in very good agreement with
those of \cite{1991ApJ...370..604P} who use a different set of equations.

\subsection{Evolution of the system}

\begin{figure*}[t!]
  \centering
  \includegraphics[width=.48\textwidth]{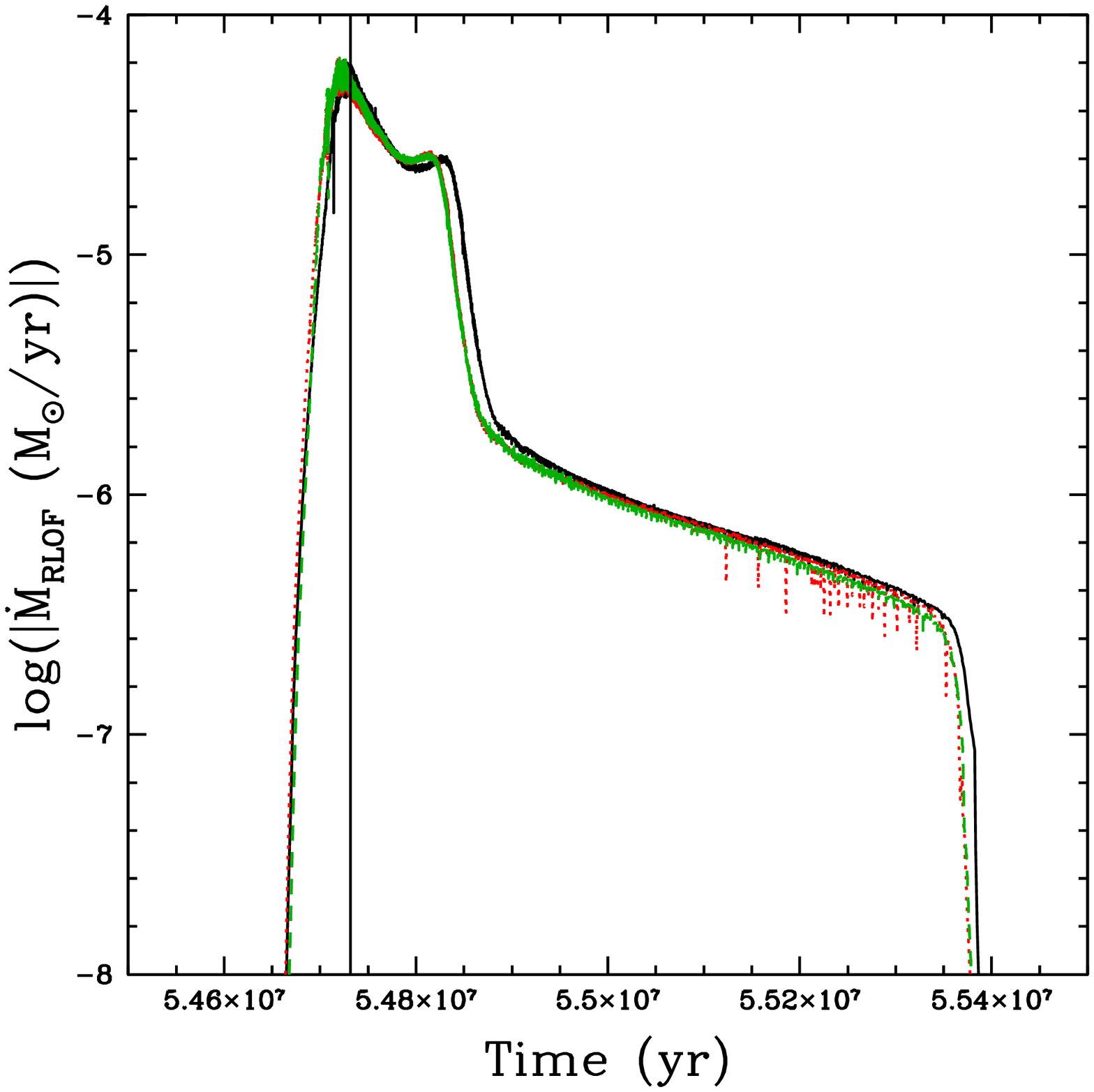}     %tl
  \includegraphics[width=.48\textwidth]{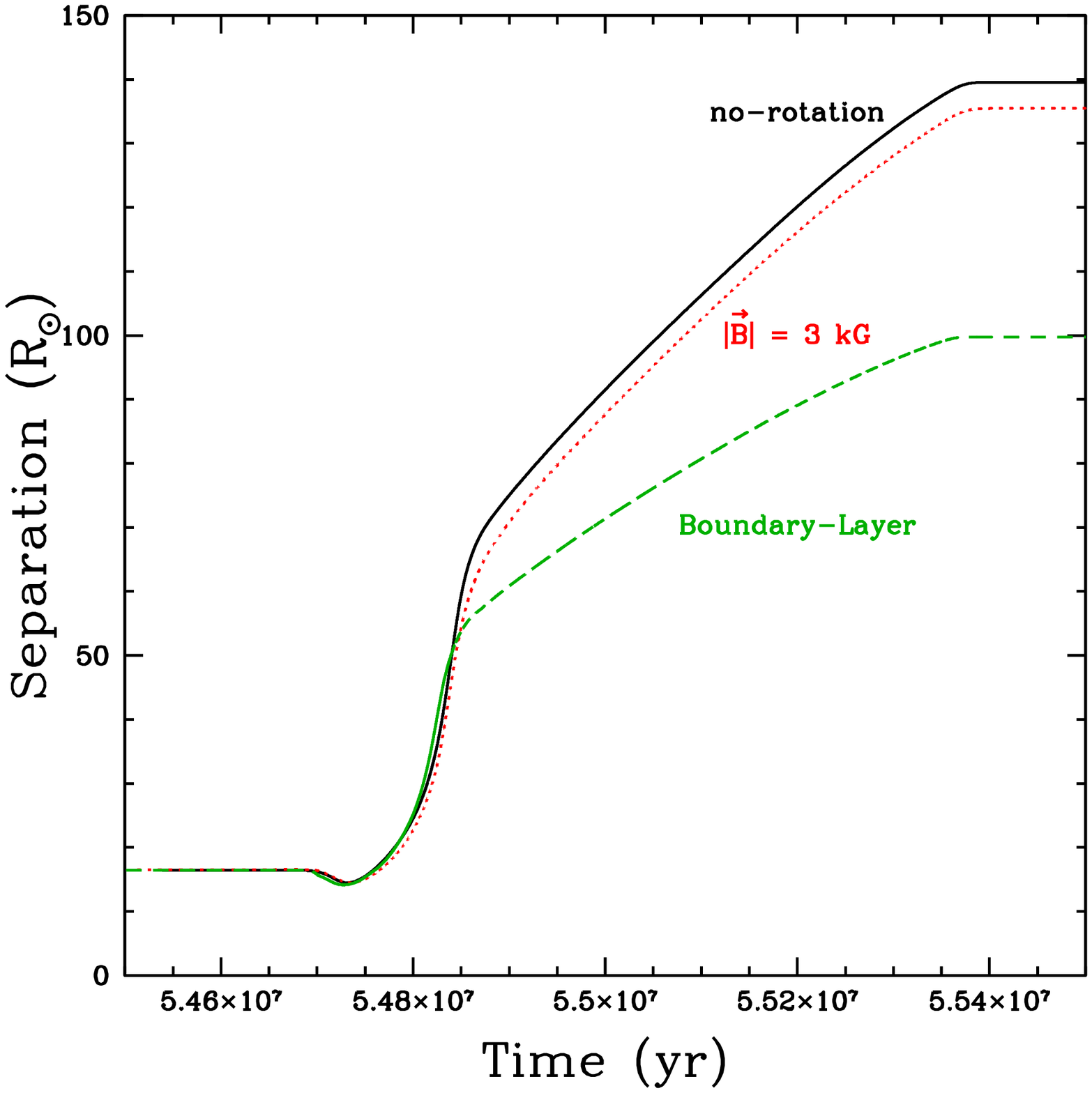}   %tr
  \includegraphics[width=.48\textwidth]{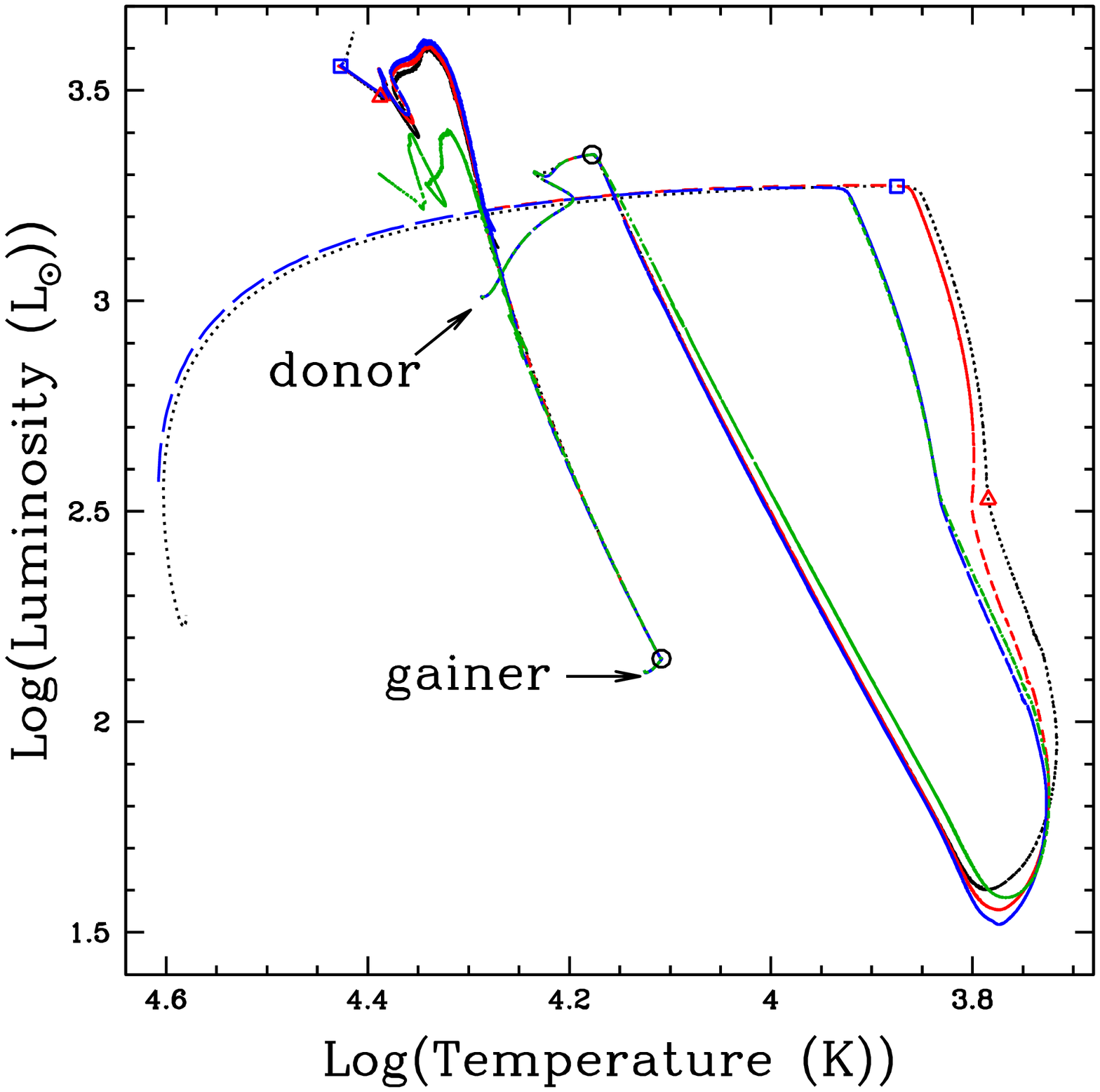}         %bl
  \includegraphics[width=.48\textwidth]{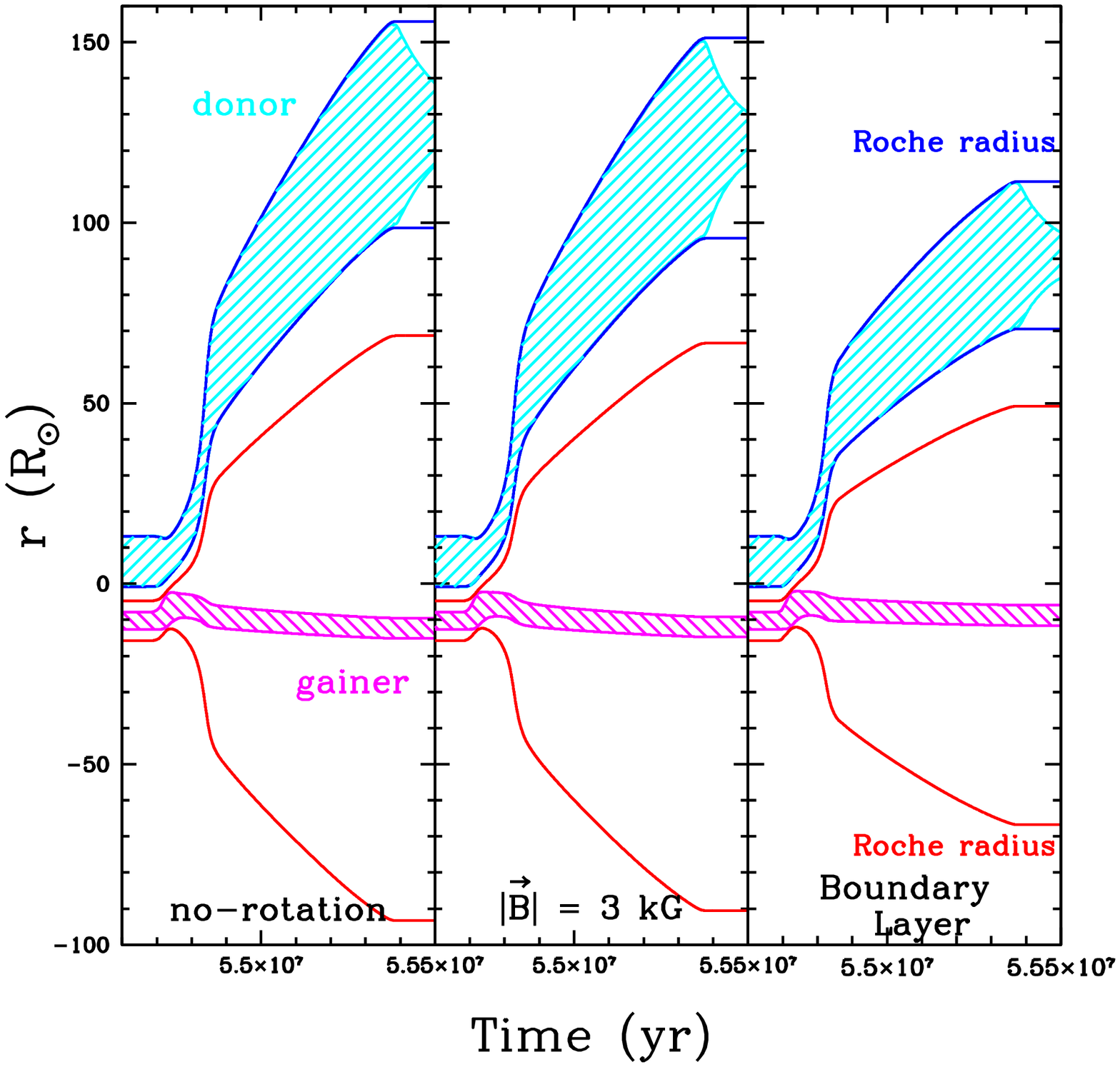}     %br
  \caption{Evolution of our 6 + 3.6~M$_{\odot}$ system with different
    braking mechanisms: solid black line: rotation-free model; dotted red
    line: magnetic field 3~kG; dashed green line: boundary-layer
    treatment. {\bf Top~Left:} Mass accretion rate on the gainer. The
    vertical line represents the epoch when the mass ratio $q$~=~1. {\bf
      Top~Right:} Orbital separation. {\bf Bottom~Left:} HR diagram for the
    three configurations: solid black: rotation-free model; dotted red:
    3~kG magnetic field; dashed blue: boundary-layer treatment; dot-dashed
    green: hotspot (see Appendix~\ref{ap:hotspot}). The open circles
    correspond to the beginning of mass transfer, triangles to the
    transition between the rapid and quiescent phase of mass transfer, and
    squares to the end of mass transfer (only on the rotation-free
    model). {\bf Bottom~Right:} Extension of the donor (top cyan hatched
    area) and gainer (bottom magenta hatched area) within their Roche lobes
    $R_{\mathrm{L}}$ (donor: solid blue lines; gainer: solid red lines) for
    the three cases. The $y$-axis represents the position relative to the
    system's center of mass. The gap between the Roche lobe radii
    $R_{\mathrm{L}}$ of the two stars is due to the fact that
    $R_{\mathrm{L}}$ is defined as the radius of a sphere with a volume
    equivalent to the Roche lobe and therefore differs from the true
    location of the $\mathcal{L}_1$ point.}
  \label{fig:general_evol}
\end{figure*}

The evolution of the binary parameters strongly depends on the spin-down
mechanism because it controls how much mass and angular momentum are
transferred.

The top right panel of Fig.~\ref{fig:general_evol} shows the evolution of
the separation for the considered spin-down mechanisms. The
`boundary-layer' simulation ends up with a shorter orbital separation
(about 100 R$_{\odot}$, orbital period of 36 days) than the magnetic-field
case (about 135 R$_{\odot}$, orbital period of 59 days) which is slightly
less than the canonical `rotation-free' model.

When a boundary layer forms, the star accretes the maximum allowed amount
of angular momentum necessary to maintain its rotational velocity at the
critical value. This process minimises the angular momentum returned to the
orbit, hence leading to a smaller orbital separation compared to the other
mechanisms. For the disc-locking mechanism, the gainer's spin is kept below
the critical rate so more angular momentum is evacuated leading to longer
period systems. For the stronger wind-braking mechanism (as compared to
disc-locking), more angular momentum will be lost from the system,
therefore reducing the orbital separation.

The differences in the mass-transfer rates between the three aforementioned
models are small (see Fig.~\ref{fig:general_evol}, top left panel) since
the separation does not vary between the considered models during the rapid
phase of mass transfer. Therefore, the Roche radius, and in turn the
mass-transfer rate, of the donor remain unchanged. Then, during the
quiescent phase, differences in the separation are negligible on the (weak)
mass-transfer rate. As a result, the final masses are exactly the same. The
donor star ends-up with a mass of 0.8~M$_{\odot}$ and the gainer at about
8.75~M$_{\odot}$. The final mass ratio is $q$~=~0.096 for the star-disc
spin-down mechanism and $q$~=~0.098 for the wind-braking mechanism.

During the mass-transfer phase, the donor radius equals the Roche radius,
which depends on the separation and thus on the braking mechanism. As the
Roche radius is smaller for the boundary-layer model than for the
magnetic-braking mechanism, so is the donor's radius (by about 30\%;
bottom-right panel of Fig.~\ref{fig:general_evol}).

The bottom-left panel of Fig.~\ref{fig:general_evol} presents the
Hertzsprung-Russell diagram (HRD) for the same three configurations. There
are no differences between these models for the gainer star since roughly
the same amount of matter is accreted: the radiative accretor ascends the
main sequence at a slightly higher luminosity. On the other hand, the
donor's evolution exhibits small differences during the quiescent
mass-transfer phase due to the different evolutionary histories of the
separation and stellar radius. By the end of the simulation all tracks
converge to the same point.

\subsection{Hotspot}
\label{sec:hotspot_res}

\begin{figure*}[t!]
  \centering
  \includegraphics[width=.48\textwidth]{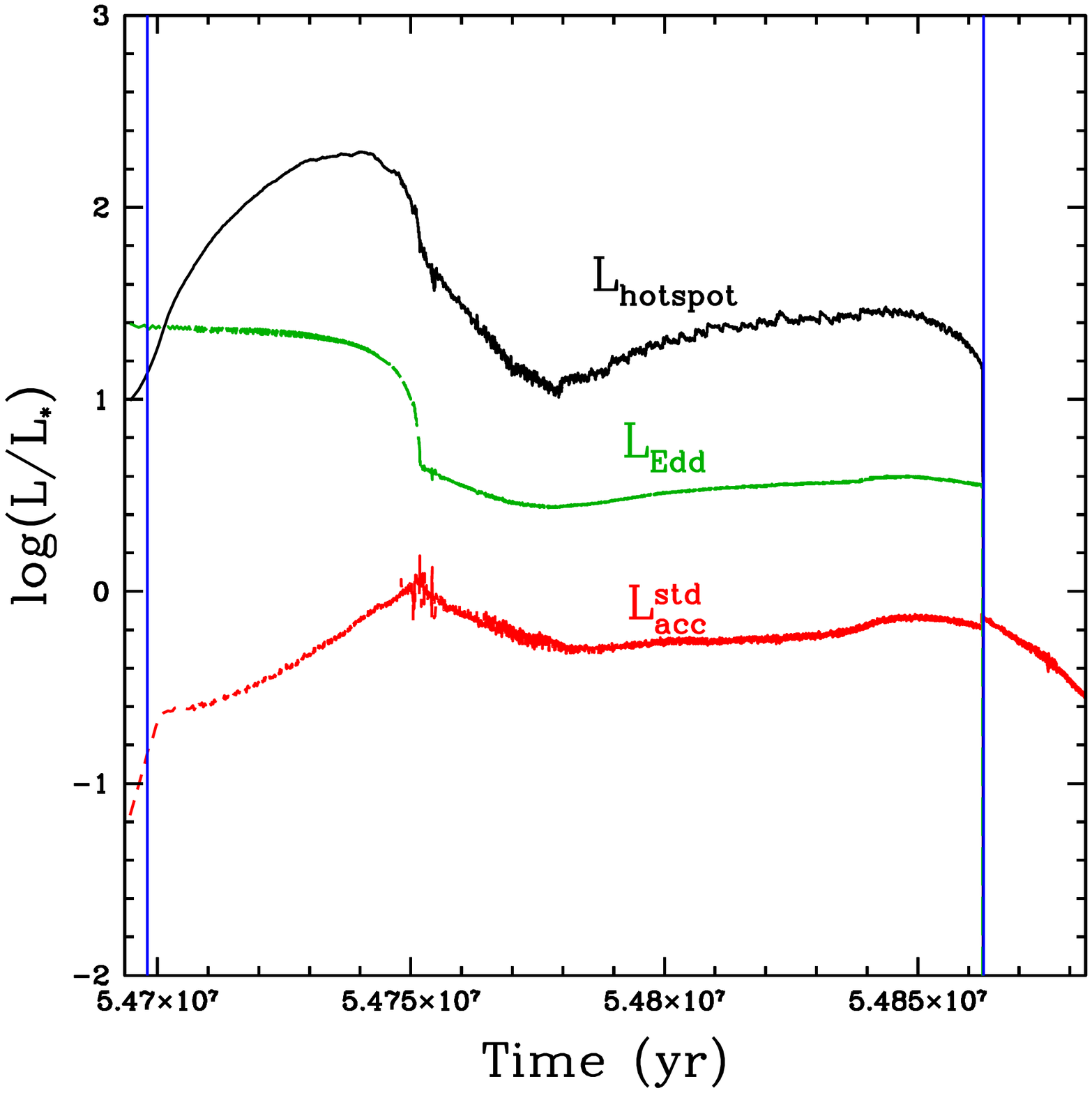}  %tl
  \includegraphics[width=.48\textwidth]{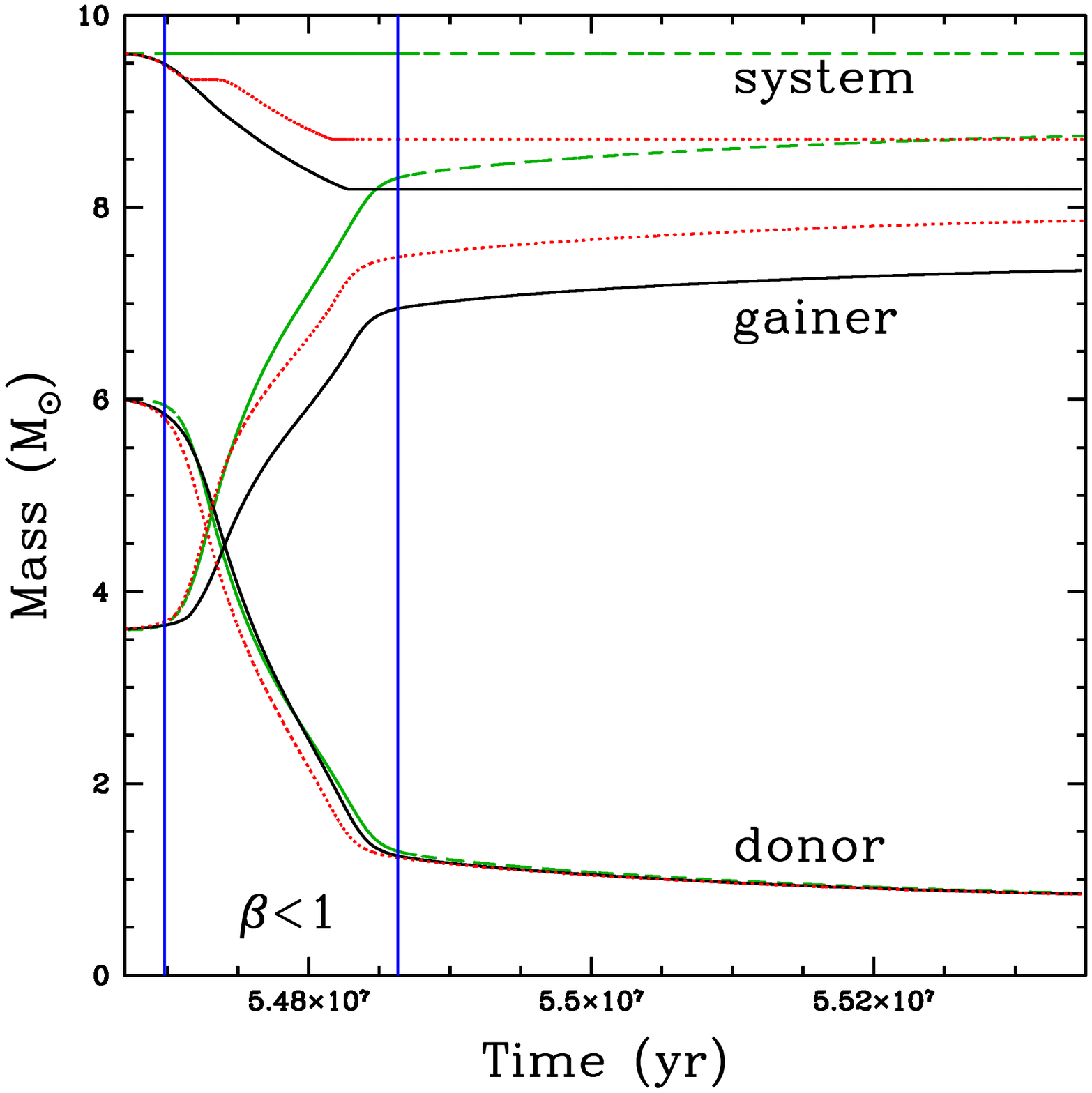}     %tr
  \includegraphics[width=.48\textwidth]{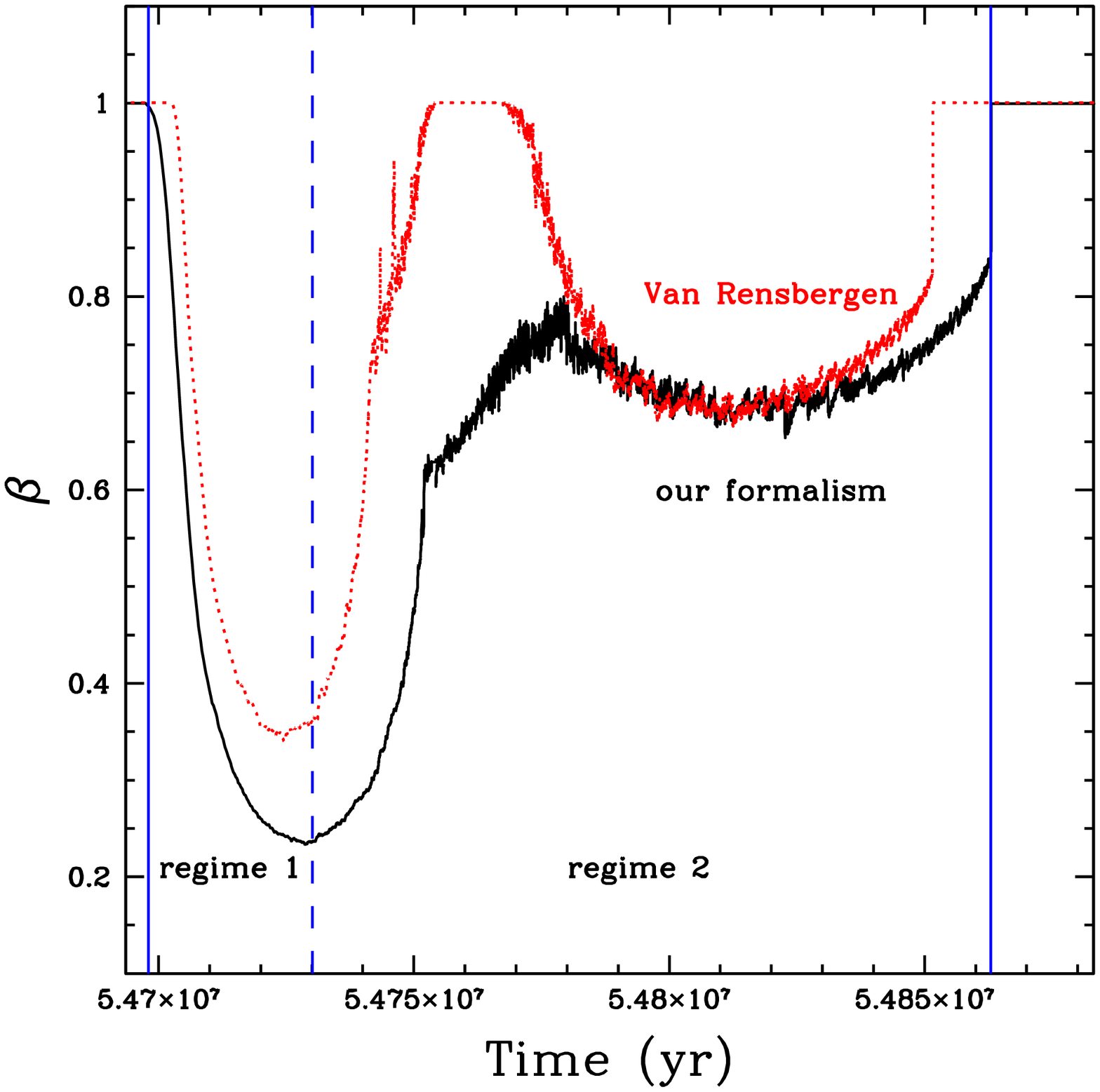}  %bl
  \includegraphics[width=.48\textwidth]{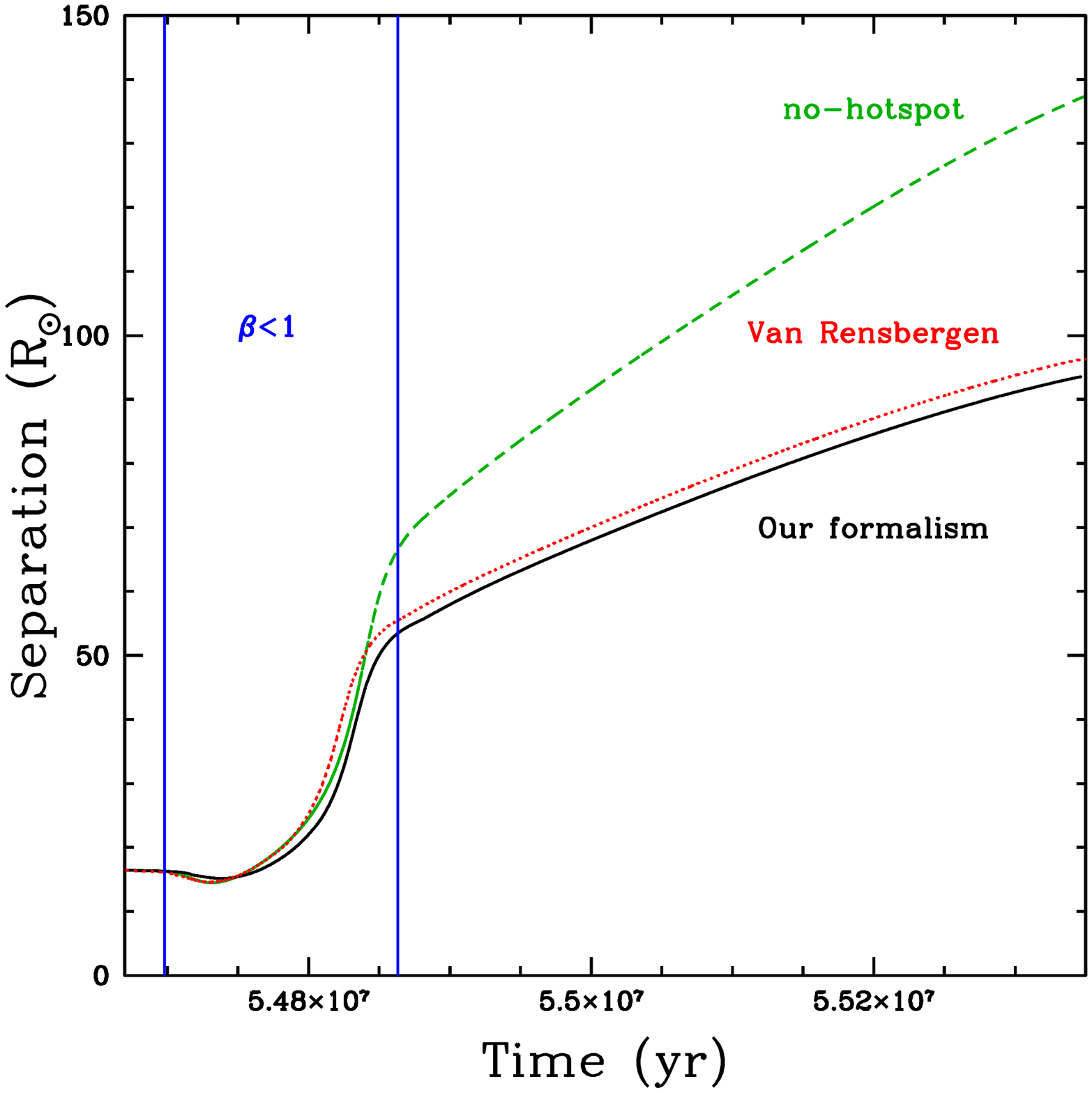}   %br
  \caption{Evolution of our 6 + 3.6~M$_{\odot}$ system with hotspot
    formation. {\bf top~left:} normalised hotspot $L_{\mathrm{hotspot}}$
    (solid black line), accretion $L_{\mathrm{acc}}^{\mathrm{std}}$ (dashed
    red) and Eddington (dotted green) luminosities. {\bf top~right:}
    evolution of the masses with (solid black) and without (dashed blue) our
    hotspot formalism compared to \cite{2008A&A...487.1129V} hotspot
    formalism (dotted red). {\bf bottom~left:} evolution of $\beta$ for
    the gainer using our (solid black) and \cite{2008A&A...487.1129V}
    hotspot formalism (dotted red). {\bf bottom~right:} evolution of the
    orbital separation (same colours as top right panel).}
  \label{fig:hotspot}
\end{figure*}

The hotspot formalism derived in Appendix~\ref{ap:hotspot} provides a
self-consistent determination of the $\beta$ parameter. In this section, it
is applied in conjunction with the boundary-layer formalism to avoid
super-critical velocities. During the rapid phase, the hotspot luminosity
$L_{\mathrm{hotspot}}$ (Fig.~\ref{fig:hotspot}, top left) is higher than
the critical Eddington value $\tilde{L}_{\mathrm{edd,g}}$
(Eq.~(\ref{eq:Ltilde})) and largely exceeds the accretion luminosity
\begin{equation}
  \label{eq:std_lacc}
  L_{\mathrm{acc}}^{\mathrm{std}} = \frac{G
    M_{\mathrm{g}} \dot{M}_{\mathrm{acc,g}}^{\mathrm{RLOF}}}{R_{g}} ,
\end{equation}
because of the small impact area (Eq.~(\ref{eq:LHS})). This hotspot
formalism leads to a non conservative evolution and small values of $\beta$
during both the rapid and quiescent phases (see bottom-left panel of
Fig.~\ref{fig:hotspot}). While almost the same amount of mass is lost by
the donor\footnote{Changes in the mass-accretion rate affect the mass-loss
  rate of the donor star because the modification of the mass ratio impacts
  on the Roche radius.} (top-right panel of Fig.~\ref{fig:hotspot}), the
gainer's final mass is quite different: 7.4~M$_{\odot}$ versus
8.8~M$_{\odot}$ for the fully conservative case. In total 1.4~M$_{\odot}$
are expelled from the system but this barely affects the final mass ratio;
0.1 for the conservative model compared to 0.12 for the hotspot model.

There are two regimes for the evolution of $\beta$ (see
Fig.~\ref{fig:hotspot}, bottom left). The first one occurs during the onset
of mass transfer when the gainer is still slowly rotating
($\Omega_{*}/\Omega_{\mathrm{crit}} \la 0.2$). This phase is characterised
by a drop in $\beta$ as the first term of Eq.~(\ref{eq:Macc_crit}),
$(\tilde{L}_{\mathrm{edd,g}} - L_{\mathrm{g}})/\tilde{K}$, dominates over
the second (rotational) term, $\Delta E_{\mathrm{rot}} / \Delta t =
(E_{\mathrm{rot}}^{\prime} - E_{\mathrm{rot}})/\Delta t$. With increasing
velocity, the star enters the second regime where the rotational term,
which is proportional to $\Omega_{*}$, becomes dominant. During this phase
$\beta$ starts to increase again because
$\dot{M}_{\mathrm{acc}}^{\mathrm{crit}}$ gets higher. Eventually, the star
reaches its critical spin velocity and $\beta$ saturates around
0.6-0.7. The noise in the $\beta$ profile is due to the explicit dependence
of the rotational term on the time-step $\Delta t$.

Once the mass ratio has been reversed, the orbital separation starts to
increase, leaving space for the formation of an accretion disc. At time $t
\approx 5.486\times 10^7$yr, the stream no longer impacts the star and our
hotspot formalism stops applying, terminating the non-conservative
evolution.

In the Hertzsprung-Russell diagram (Fig.~\ref{fig:general_evol},
bottom-left panel), the gainer is less luminous because it accretes less
mass. For the donor, on the other hand, the presence of a hotspot does not
alter its evolution in the HRD. The main effects of this non-conservative
evolution are the slight decrease in the final orbital separation (94
R$_{\odot}$ vs 100 R$_{\odot}$ in the conservative case) and a slower
acceleration of the gainer's rotational velocity.

The parameter $\tilde{K}$ defining the hotspot luminosity
(Eq.~(\ref{eq:L_hotspot})) has a strong impact on the critical
mass-accretion rate and thus on $\beta$. Given the range of values for
$\tilde{K}$ derived from observations (from $\tilde{K} \approx 1$ to
$\approx 700$; \citealt{2011A&A...528A..16V}), a fully conservative or a
significantly non-conservative evolution can result from the
simulations. In our calculations, $\tilde{K}$ varies between ~125 and ~161
and the total amount of mass ejected is ~1.4~M$_{\odot}$. If we were to use
$\tilde{K} = 1$, we would have a fully conservative evolution for the same
system.

The opacity of the material at the impact location (where the Eddington
luminosity is computed) is set to the photospheric value that is higher
than the Thomson electron-scattering opacity $\kappa_{\mathrm{es}}$
generally used. Given the sensitivity of $\tilde{L}_{\mathrm{edd,g}}$ to
this quantity, we ran a simulation using $\kappa_{\mathrm{es}}$ for the
computation of $\tilde{L}_{\mathrm{edd,g}}$ and found a fully conservative
evolution. The opacity has thus a strong impact on $\beta$. It is not
straightforward (and beyond the scope of this article) to precisely compute
the physics (opacity, shock formation) at the impact location, since this
depends on the temperature of the disturbed material and the penetration
depth of the stream. This depth can be estimated by equating the stream ram
pressure $\rho \vert v^{2} \vert$ to the stellar gas pressure
\citep{1976ApJ...206..509U}. However, complications arise from the fact
that the rotation of the star makes the stream impact upon a new
undisturbed part of the star at each instant. Moreover, the matter can be
ejected directly from the impact location but also from its trail (i.e.,
from the perturbed material no longer at the impact location) where the
opacity may be different.

Due to the large amount of mass transferred in massive short-period
binaries, the gainer's radius may increase, until it fills its own Roche
radius and the system may then evolve into contact. However, the presence
of a hotspot can prevent such evolution because less mass is accreted onto
the gainer star.

In contrast to our simulation, the model of \cite{2008A&A...487.1129V} for
the same system (with initial masses 6 + 3.6~M$_{\odot}$ and initial period
P$_{\mathrm{init}}$ = 2.5 d) is fully conservative (neglecting winds).  It
is not straightforward to disentangle the differences between the two runs,
because of the different approaches used to compute various physical
quantities. For instance, the stream velocity is computed in
\textsc{Binstar} via ballistic motion while \cite{2008A&A...487.1129V} use
geometrical arguments. They also use a parametric tidal spin-down
prescription which is more efficient than the one computed in
\textsc{Binstar}, derived from the stellar structure. Therefore, in their
models, the gainer star is more efficiently spun down and in turn has a
lower mass-loss rate due to the higher $\tilde{L}_{\mathrm{edd,g}}$
(Eq.~(\ref{eq:Ltilde2})). The computation of the opacity also differs; we
use the photospheric opacity (at $\tau = 2/3$) while they use a parametric
expression based on the electron-scattering opacity. For comparison, we
implemented their formulation in \textsc{Binstar} and found that less mass
is lost from the system ($\approx$0.9~M$_{\odot}$) mainly because they
neglect the term $a_{\mathrm{g}}^{2} \Omega_{\mathrm{orb}}^{2}$ in
Eq.~(\ref{eq:Macc_crit}) which is not negligible ($a_{\mathrm{g}}^{2}
\Omega_{\mathrm{orb}}^{2}/v^{2} \approx 0.2$) and contributes to further
decrease $\dot{M}_{\mathrm{acc,g}}^{\mathrm{crit}}$. As a result, the final
separation (bottom-right panel of Fig.~\ref{fig:hotspot}) also changes
between the two hotspot formalisms since it is linked with the masses of
the two stars. Our less conservative model leads to a shorter-period
system. Nevertheless, both formalisms show the same two phases for the
evolution of $\beta$.

\subsection{Hotspot and magnetic braking}
\label{sec:results:betaB}

\begin{figure*}[t!]
  \centering
  \includegraphics[width=.48\textwidth]{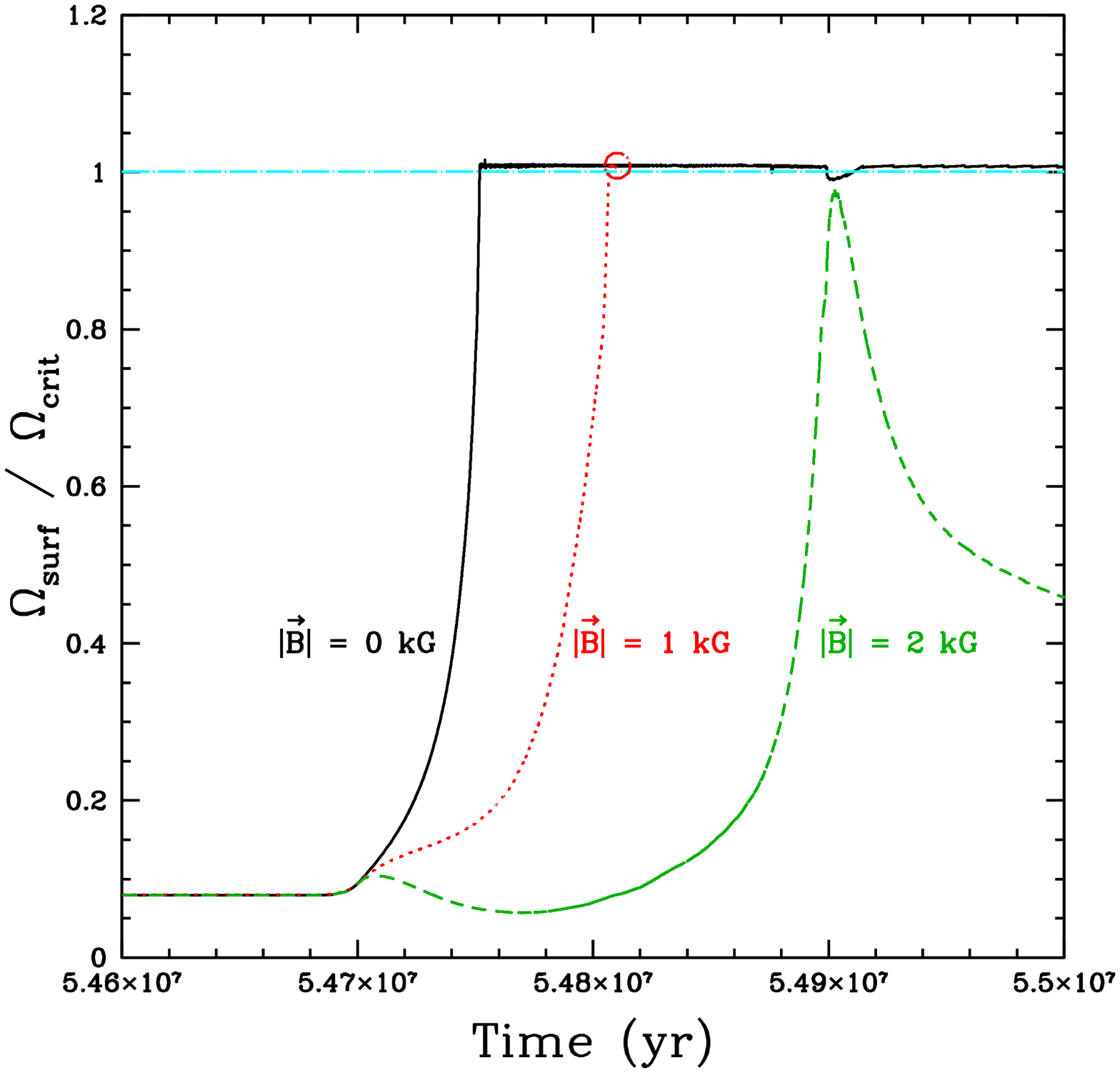} %tr
  \includegraphics[width=.48\textwidth]{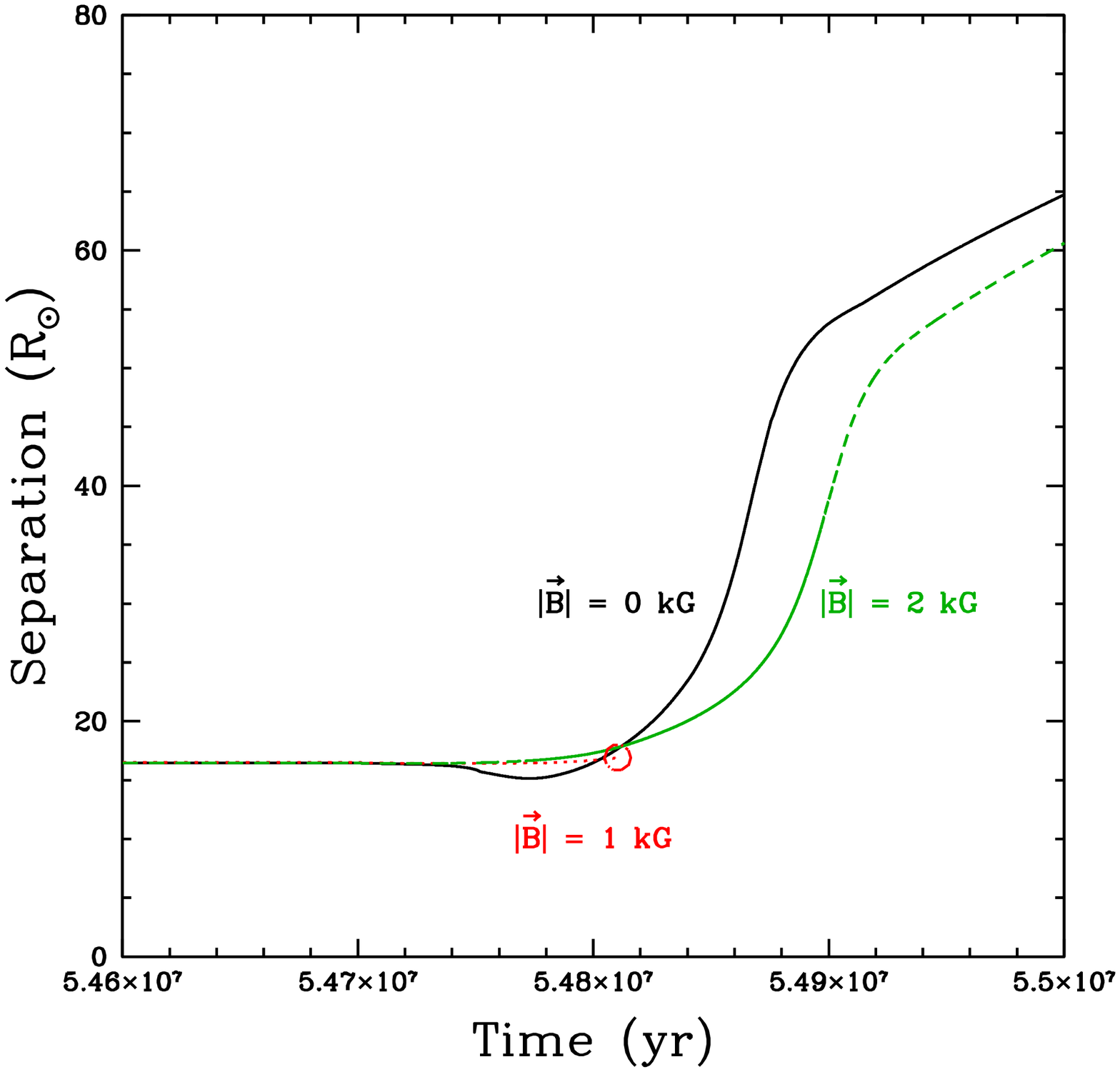}   %bl
  \includegraphics[width=.48\textwidth]{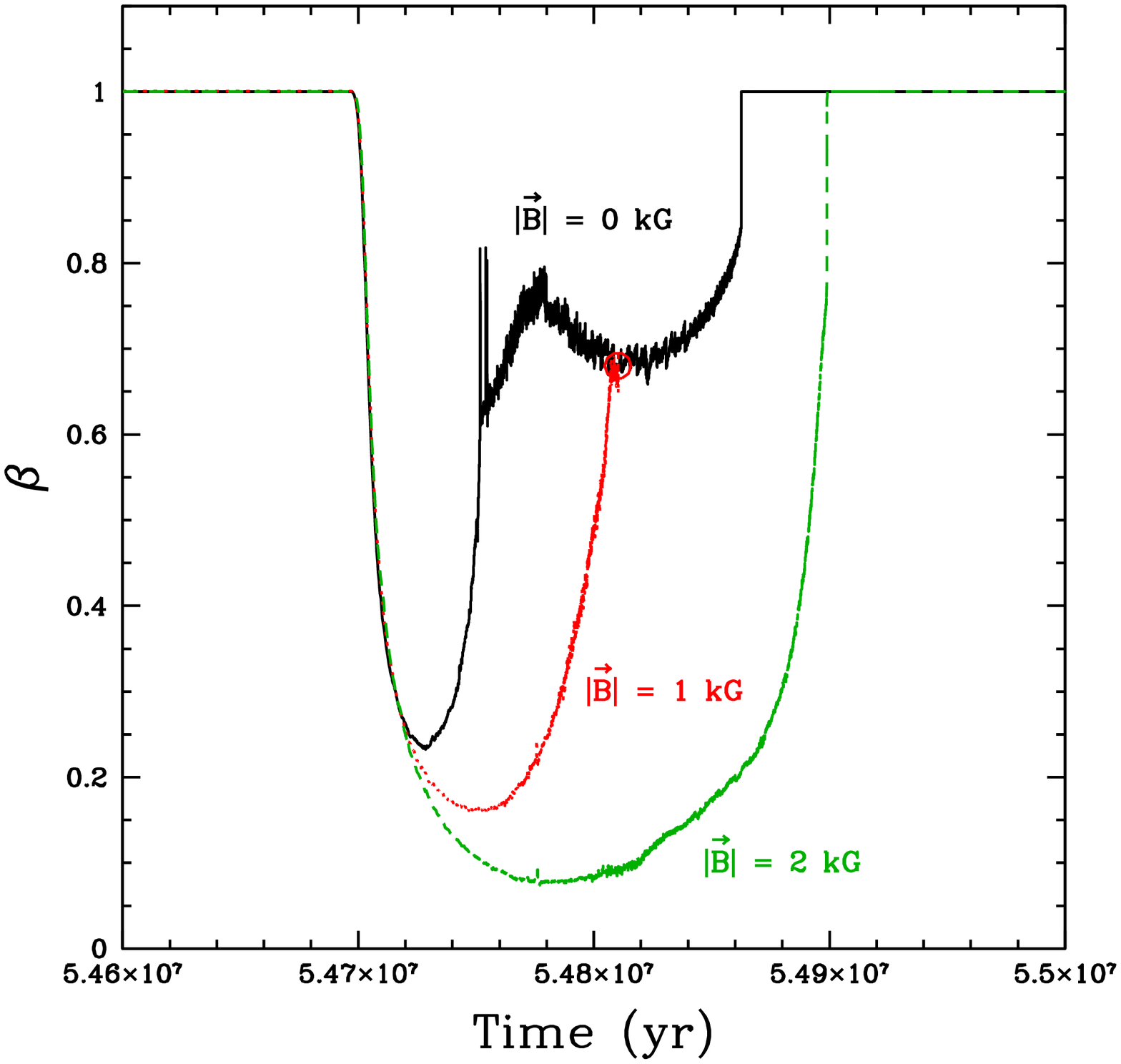}  %tl
  \includegraphics[width=.48\textwidth]{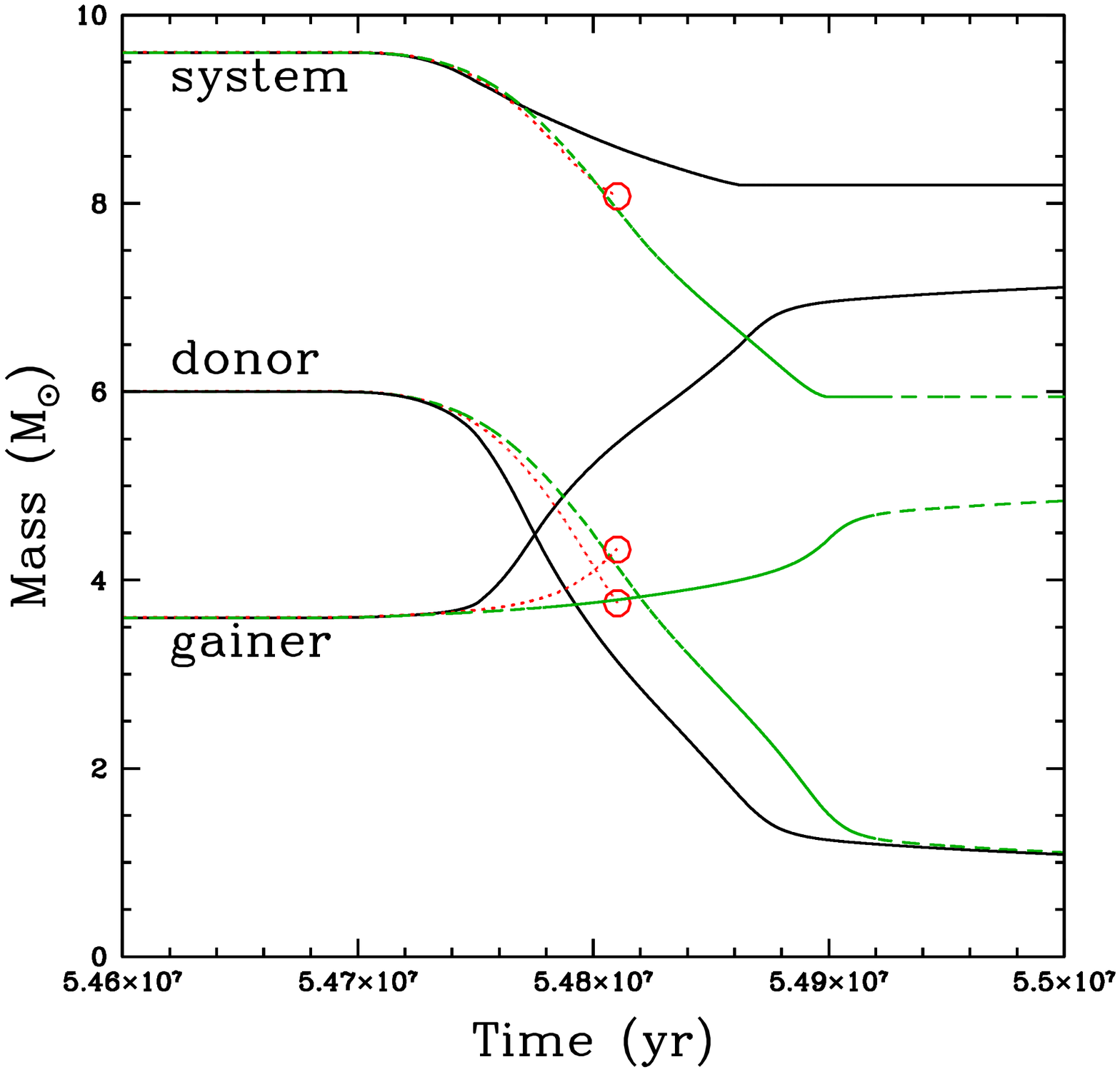}  %br
  \caption{Evolution of our 6 + 3.6~M$_{\odot}$ system with
    P$_{init}$~=~2.5~days with hotspot and different magnetic field
    strengths. Black: hotspot with boundary layer (BL) and no magnetic
    field. Red: hotspot (no BL) + magnetic field $\vert
    \vec{B}_{*}\vert$~=~1~kG. Green: hotspot (no BL) + magnetic field
    $\vert \vec{B}_{*}\vert$~=~2~kG. The red simulation stops when the
    gainer reaches its critical spin-angular momentum (indicated by a red
    circle). {\bf Top Left:} Surface spin-angular velocity. {\bf Top
      Right:} Orbital separation. {\bf Bottom Left:} Accretion efficiency
    $\beta$. {\bf Bottom Right:} Masses.}
  \label{fig:betaB}
\end{figure*}

We mentioned that $\beta$ is a key parameter for the magnetic-wind
braking. In Sect.~\ref{sec:RES:mag}, only stellar winds have been
considered as a mass-loss mechanism from the system but the matter
emanating from the hotspot might contribute to the magnetic-wind braking by
increasing the mass-loss rate $\dot{M}_{\mathrm{W}}$ in
Eq.~(\ref{eq:mdot_wind}). As a result, a lower magnetic-field strength is
needed to keep the gainer's spin velocity below the critical value. 

In the configuration with a hotspot plus a 1~kG magnetic field and no
boundary-layer mechanism, the gainer reaches the critical rate,
although later than in the conservative case (Fig.~\ref{fig:betaB}). Above
2~kG, the critical rotation can be avoided but the required field strength
still remains higher than measured in Algols.

As expected, coupling the expelled material to the magnetic field will
increase the system angular momentum loss and thus the separation. As a
consequence, the impact velocity will be greater and according to
Eq.~(\ref{eq:Macc_crit}), the threshold for mass ejection
$\dot{M}_{\mathrm{acc}}^{\mathrm{crit}}$ will be lower. As explained
previously, the value of $\beta$ drops until the rotational term
$\frac{E_{\mathrm{rot}}^{\prime}-E_{\mathrm{rot}}}{\Delta t}$ starts to
dominate. The $\beta$ profiles further diverge as time goes on because of
the increasing difference in the gainer's luminosities due to their
different masses. In the case of a strong magnetic field (2~kG), $\beta$
drops to lower values ($\beta \approx 0.1$) and for a much longer
duration. The donor reaches the same final mass ($M_{\mathrm{d}} \approx
0.8$~M$_{\odot}$) but because 3.6~M$_{\odot}$ are removed from the system,
the final mass ratio increases significantly from 0.1 to 0.25.

In their study, \cite{2010MNRAS.406.1071D} use a constant value of $\beta$
ranging from 0.1 to 0.9 to account for the system mass-loss. This is in
contrast to our approach where $\beta$ is computed consistently. Our
simulations cannot maintain values as low as those used by these authors
during the entire phase of mass transfer and this explains why in our case
the magnetic-wind braking is not as efficient. 
Because of our poor knowledge of the hotspot outflow geometry and its link
with the magnetic field, that coupling should be regarded as exploratory.

\section{Can observations help constraining Algol evolution?}
\label{sec:discussions}

\subsection{Predicting the mass ejected from the hotspot}
\label{sec:DIS:mass_loss}

\begin{figure}[t!]
  \centering
  \includegraphics[height=.48\textwidth,angle=-90]{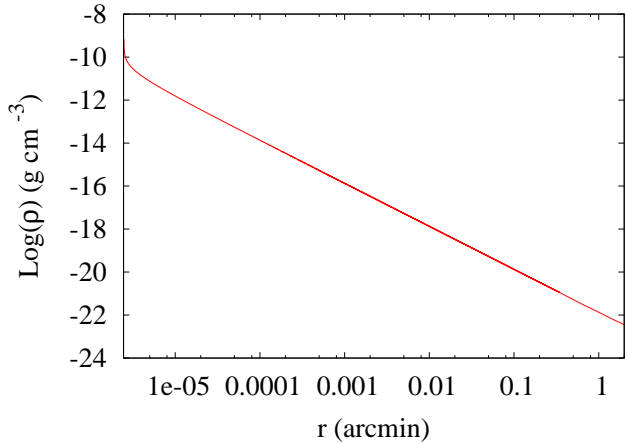}
  \caption{Density profile of the matter expelled by the hotspot and
    surrounding the system. The $x$-axis represents the angular extent
    for a system located at 100~pc. The profile starts at the stellar
    radius of the gainer ($R_{*}$~=~2.2$\times$10$^{11}$~cm) and is cut at
    2~arcmin. The density drops below the mean density of the ISM value at
    12~arcmin from the star, i.e. 1.05$\times$10$^{18}$~cm.}
  \label{fig:rho}
\end{figure}

It is not clear what happens to the matter ejected via the hotspot, whether
it freely escapes the system as assumed in our simulation or if it remains
in the gainer's Roche lobe to be eventually re-accreted. Since in our model
the hotspot luminosity exceeds the Eddington luminosity, we expect the
matter to reach the escape velocity
\begin{equation}
  v_{\mathrm{esc}}(R_{*}) = \sqrt{\frac{2 G M_{\mathrm{eff}}}{R_{*}}} ,
\end{equation}
where $M_{\mathrm{eff}}$ is the effective gravitational mass felt by the
wind \citep{1975ApJ...195..157C}
\begin{equation}
  M_{\mathrm{eff}} = M_{*} (1-\Gamma_{\mathrm{L}}) = M_{*} \left(1 -
    \frac{\kappa_{es} L_\Sigma}{4\pi c G M_{*}}\right) ,
\end{equation}
where $\kappa_{es}$ is the Thomson electron-scattering opacity. In this
equation, the luminosity $L_\Sigma$ is due to both the star and the
hotspot. We further assume that, once ejected, the flow is accelerated by
radiation pressure. Following \cite{1975ApJ...195..157C} prescription for
radiation-driven winds, the material velocity at a distance $r$ from the
gainer is then given by
\begin{equation}
  v(r,t) = \sqrt{\frac{\alpha}{1-\alpha}}\,v_{\mathrm{esc}}(R_{*}(t))
  \sqrt{1-\frac{R_{*}(t)}{r}} ,
\end{equation}
where $\alpha$ is a force multiplier parameter \citep{1999isw..book.....L}
which has been set to three different values: $\alpha = 0.465$ (for
$T_\Sigma = $6$\times$10$^{3}$~K\footnote{$T_\Sigma$ is the equivalent
  temperature of the hotspot plus the star, obtained by summing the fluxes
  and applying Stefan-Boltzmann's law.}); $\alpha = 0.5$ (for $T_\Sigma =
$3$\times$10$^{4}$~K typical of hotspot temperatures;
\citealt{2010A&A...510A..13V}); $\alpha = 0.640$ (for $T_\Sigma =
$5$\times$10$^{4}$~K). For the sake of simplicity, the influence of the
secondary star and the internal energy of the flow have been neglected.

The density of the material surrounding the system at a given time $t$ and
distance $r$ is calculated using the continuity equation
\begin{equation}
  \rho(r,t) = \frac{\dot {M}_{\mathrm{flow}}(r,t)}{\tilde{\sigma} r^{2} v(r,t)} ,
\end{equation}
where $\tilde{\sigma}$ is some solid angle over which the material is
ejected. The geometry of the mass ejection is badly constrained but will
probably be collimated due to the small area of the hotspot and for
simplicity, we use $\tilde{\sigma} =\pi$. The mass flow
$\dot{M}_{\mathrm{flow}}(r,t)$ at time $t$ and position $r$ is equal to the
system mass-loss rate $(1 - \beta)
\dot{M}_{\mathrm{loss,d}}^{\mathrm{RLOF}}(t)$ estimated at the time such
that $r= \int_0^t v(r,\tau)\,\mathrm{d}\tau$. In other words, $r$ is the
distance travelled by the ejected layer since the beginning of the
non-conservative phase.

In this simplistic model it is also possible to follow the chemical
stratification of the expelled material because we know from the stellar
model the composition of the ejected layers at each time-step. The chemical
changes are small until the gainer's mass has dropped below $\approx
3M_\odot$ and the CNO-processed layers start to be ejected. By the end of
the non-conservative evolution, the abundance of $^{14}$N has increased by
a factor of 5 while $^{12}$C has been almost completely depleted. Since
C+N+O remains constant, the global metallicity of the ejected material is
almost constant.

The final density profile (when the non-conservative phase ends) is shown
in Fig.~\ref{fig:rho}. The $x$-axis represents the angle subtended by the
flow for a system located at 100~pc (to fix the ideas, we remind that
$\beta$ Persei is situated at $\approx$30~pc and $\beta$~Lyr\ae{} at
$\approx$300~pc). The density drops below the mean interstellar medium
(ISM) density ($\approx 10^{-24}$~g~cm$^{-3}$,
\citealt{2001RvMP...73.1031F})\footnote{In principle, it is the equilibrium
  between the ram pressures of the wind and of the ISM that imposes the
  boundary of the stellar flow. However, since the hotspot stream velocity
  is much larger than the ISM velocity -- a few tens of km~s$^{-1}$ at most
  --, that boundary is even further away than the one based on the density
  used here.} at 12~arcmin from the star,
i.e. 1.05$\times$10$^{18}$~cm. Closer to the star, the density rises up to
10$^{-9}$ g cm$^{-3}$ in the thin region of the wind acceleration. Finally,
we report that the parameter $\alpha$ has almost no impact on the density
profile in the inner region (inside the 2 arcmin region).

The calculation of the radiative transfer through this surrounding material
would allow a meaningful comparison with infrared flux observations, but it
is beyond the scope of this article and deferred to a future work.

\subsection{Observational constraints on the spin-down mechanism}
\label{sec:RES:mass_loss}

In this section, we attempt to identify observational constraints that can
corroborate or invalidate spin-down mechanisms and non-conservative
evolution.

\subsubsection{Classification of Algol systems}
\label{sec:RES:classification}

\begin{figure}[t!]
  \centering
  \includegraphics[width=.48\textwidth]{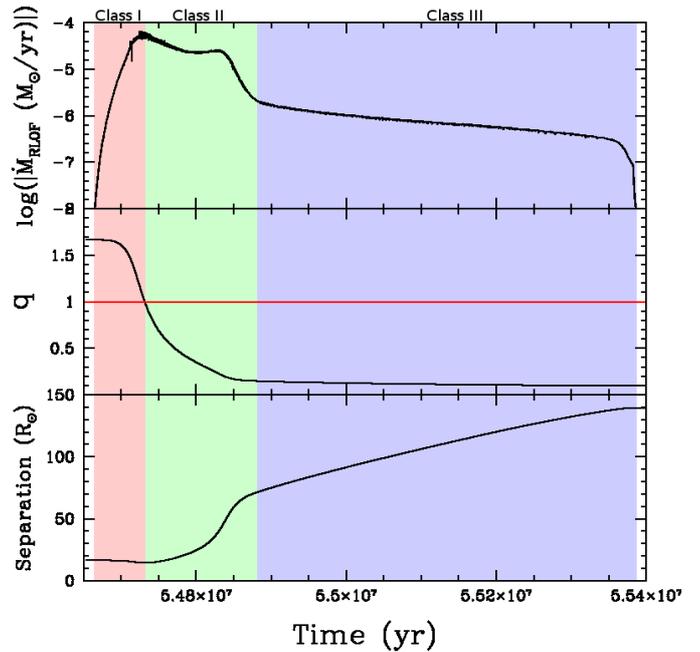}
  \caption{Evolution of the mass-transfer rate (top-panel), mass ratio
    ($q=M_{\mathrm{d}}/M_{\mathrm{g}}$) and separation for our 6 +
    3.6~M$_{\odot}$ system with P$_{\mathrm{init}}$~=~2.5~days. The different
    evolutionary states are coloured in red (Class~I), green (Class~II) and
    blue (Class~III).}
  \label{fig:prototype}
\end{figure}

\begin{table*}
  \begin{center}
    \caption{Observational prototypes for the different classes introduced
      in Sect.~\ref{sec:RES:classification} and
      Fig.~\ref{fig:prototype}. Sp$_{\mathrm{d}}$ (Sp$_{\mathrm{g}}$) is
      the spectral type of the donor (gainer). The gainer of W~Ser is
      embedded in an accretion disc and its spectral type is unknown but
      believed to be B-A.}
    \begin{tabular}{cccccccc}
      \hline
      \hline \\
      Theoretical class & Prototype & M$_{\mathrm{d}}$ + M$_{\mathrm{g}}$
      (M$_{\odot}$) & $q$ = M$_{\mathrm{d}}$ / M$_{\mathrm{g}}$ &
      $\dot{M}_{\mathrm{RLOF}}$ (M$_{\odot}$ yr$^{-1}$) &
      Period (d) & Sp$_{\mathrm{d}}$ + Sp$_{\mathrm{g}}$ & Ref.\\ [1em]
      \hline
      \hline \\
      Class~I  & SV~Cen      & 8.56 + 6.05 & 1.41 & 1.626$\times$10$^{-4}$ &
      1.6585 & B1 + B4.5 & 1,2,4\\
      & UX~Mon      & 3.90 + 3.38 & 1.15 &5.46$\times$10$^{-6}$  &
      5.904  & A7p + G2p & 3\\ [1em]
      \hline \\
      Class~II & $\beta$~Lyr & 4.25 + 14.1 & 0.30 & 3.440$\times$10$^{-5}$ &
      12.9138 & B6-B8 II, + B0.5 V & 1,2,5\\
      & W~Ser       & 0.970 + 1.510 & 0.64 & $\sim$1$\times$10$^{-7}$ &
      14.154 & F5III + B-A (emb.)& 6,8 \\ [1em]
      \hline \\
      Class~III& $\beta$~Per & 0.81 + 3.7 & 0.21 &$\sim$1$\times$10$^{-11}$ &
      2.8673 & K4 + B8 & 1,7\\[1em]
      \hline
      \hline
      
    \end{tabular}
    \label{tab:prot}
    
  \end{center}
  \tablebib{(1) \citet{2011A&A...528A..16V} and references herein;
  (2) \citet{2010AIPC.1314...45V}; (3) \citet{2011A&A...528A.146S}; (4)
  \citet{1976MNRAS.176..625W}; (5) \citet{2012ApJ...750...59L}; (6)
  \citet{2004A&A...417..263B}; (7) \citet{1983ApJS...52...35G}; (8)
  \citet{2005ApJ...632..576P}: mass-transfer rate derived from
  period-change rate ($\dot{P}/P=14$~s~yr$^{-1}$) assuming a
  conservative mass transfer.}
\end{table*}

Depending on their observable properties (presence of a disc, mass ratio,
mass-transfer rate...), binary systems in a given and the same evolutionary
state from a theoretical standpoint may end up being associated with
different observational classes. Therefore, to ease the comparison between
observations and models, we define theoretical classes based on our
simulation results and associate a prototype with each class.

The first class (Class~I, see Fig.~\ref{fig:prototype}) refers to the state
before the mass-ratio reversal when the mass ratio and the orbital period
decrease. Observational prototypes are UX~Mon \citep{2011A&A...528A.146S}
or SV~Cen (see Table~\ref{tab:prot}). During this phase, mass transfer is
increasing until the system enters the rapid mass-transfer phase. These
objects can host accretion discs and exhibit bipolar jet and/or show strong
UV emission lines likely due to the formation of a hotspot
\citep{1989SSRv...50...35G}. \cite{1997AstL...23..698T} also report excess
IR emission attributed to dusty material surrounding the system.

For Class~II (Fig.~\ref{fig:prototype}), the mass ratio is reversed and the
mass-transfer rate is at a maximum. This stage is the realm of
$\beta$~Lyr\ae{} or W~Ser objects (see Table~\ref{tab:prot}). The same
features as for Class~I objects (discs, jets, hotspots, circumbinary gas)
may possibly be present but Class~II objects differ from Class~I by their
longer and increasing orbital periods. They are often referred as `active'
or `hyper-active' Algols.

The last class (Class~III, Fig.~\ref{fig:prototype}) includes quiescent
objects with their mass ratio reversed. It is during this phase that the
separation evolves the most, but the mass ratio remains roughly
constant. This class mainly corresponds to genuine Algol systems (including
the prototype $\beta$ Per; see Table~\ref{tab:prot}).

\subsubsection{Observational constraints from Class~I-II objects}
\label{sec:classI}

Spin-down mechanisms and system-mass loss may act more efficiently during
these phases (Classes~I-II) and lead to clearer observational signatures,
making these systems more interesting despite the fact that they are
short-lasting compared to Class~III objects.

The observationally-defined W~Ser systems, that we associate in our
theoretical system to Class~I-II objects, were first detected and
classified owing to their UV emission which is attributed to the impact of
the accretion stream on the edge of the accretion disc. The presence of gas
around these systems (see Sect.~\ref{sec:RES:classification}) is consistent
with mass ejection from a hotspot.

Accretion discs are frequently detected in Class~I-II systems
(\citealt{1989SSRv...50...35G,2011A&A...528A..16V}) and contribute to the
regulation of the gainer's spin velocity.  The Doppler tomography method
(mapping the circumstellar matter in a system), has been extensively
applied to $\beta$~Lyr\ae{} systems \citep{1995ApJ...438L.103R,
  1996ApJ...459L..99A, 2004AN....325..229R, 2007ApJ...656.1075M,
  2010ApJ...720..996R}. These studies reveal the presence of disc-like
structures even in the case where the gainer has a large filling factor and
where direct impact is expected instead \citep{2004ApJ...608..989B}. This
supports our paradigm of disc formation around critically-rotating gainers
and the activation of the boundary-layer mechanism (Sect.~\ref{sec:bl}) or
the disc-locking mechanism (Sect.~\ref{sec:disc_lock}), especially for
Class~I-II objects where the separation is short and the mass-transfer rate
high.

\subsection{On the rotation of Algols}
\label{sec:rot}

Most observed Algol systems have an accretor with a rotational velocity in
the range $0.1<\Omega_{*}/\Omega_{\mathrm{Kep}}<0.4$
\citep{1990AJ....100.1981V} but some of them are very rapidly rotating with
a Keplerian (or close to Keplerian) surface spin velocity
\citep{1957PASP...69..574M, 1959ApJ...130..791H, 1989SSRv...50..191W,
  1996MNRAS.283..613M}. Unfortunately, there are no observations of spin
rates for the gainer in the interesting W~Ser objects because they are
completely embedded in their accretion discs. The very high spin velocities
in genuine Algols can be regarded as the signature of recent
angular-momentum transfer or as the fact that the braking mechanism did not
have the time to operate yet and this is consistent with the boundary-layer
paradigm that necessarily leads the star to critical rotation. On the other
hand, to explain the slow rotators, another mechanism must be advocated to
spin the gainer down after the rapid phase of mass transfer. This can be
due to tides or magnetic braking (with a low magnetic field strength $<$ 3
kG), processes that were inefficient during the earlier evolution of rapid
mass transfer.

If the star is not rotating as a solid body, the outermost layers will
first be spun up and owing to the differential rotation, instabilities will
develop and allow angular momentum to be transported in the interior. In
this case, the outermost layers will be accelerated to the critical
velocity after a small amount of mass has been accreted and we may expect
the boundary-layer mechanism to be activated earlier during the
evolution. If the angular-momentum redistribution inside the star is not
efficient enough for the gainer to reach solid-body rotation by the end of
the mass-transfer phase, less angular momentum will then be stored into the
star and a larger amount given back to the orbit, thus increasing the
orbital separation.

But, rotation can also affects the mechanical and thermal
structure of the star because it brakes the spherical symmetry. Taking into
account these effects in the stellar structure equations \cite[][hereafter
referred to as `centrifugally supported models']{1976ApJ...210..184E} leads
to larger effective radii and cooler surface temperatures. As a result, and
because of the short separation, our system runs into contact just before
the reversal of the mass ratio. However this contact phase can be avoided
if the initial period is increased to $\approx$ 5 days (instead of 2.5
days). So, in order to reproduce the evolution of canonical models,
i.e. where the effects of departure from spherical symmetry are not
accounted for, our centrifugally supported models must start with a longer
initial period. It also implies that neglecting these structural effects of
rotation will produce less contact systems. Finally, we also point out that
due to the growth of the stellar equatorial radius, less room is left for
an accretion disc to form, and mass loss through the third Lagrangian point
$\mathcal{L}_{3}$ may be enhanced.

Finally, rotational instabilities also contribute to the transport of
chemicals and the mixing is likely to be very efficient in the gainer star
because of the strong torques applied at the surface
\citep{2009A&A...495..271D}. Similarly to fast rotating massive stars, we
may expect the accretor to evolve almost chemically homogeneously
\citep{1987A&A...178..159M,2010AIPC.1314..291D}, showing surface
enhancements in helium and nitrogen. These alterations of the surface
composition can decrease the envelope opacity, making the gainer star more
compact and more luminous, thus counterbalancing the radius growth
associated with the rotational deformation of the structure described
above. Clearly, stellar calculations including angular momentum transport
and rotationally induced mixing are required to clarify the situation.

\section{Conclusions}\label{sec:conclusion}

In this study, we performed the first binary-evolution calculations that
consistently take into account the torques arising from magnetic fields and
star-disc interaction. This is the first attempt to confront several
braking mechanisms with the use of the state-of-the-art binary-star
evolutionary code \textsc{Binstar}. We show that the torques due to
magnetic field and star-disc interaction can prevent the accreting star
from reaching a super-critical velocity although the magnetic-field
strength required are stronger than that of typical Algols. Tides are shown
to be inefficient, but we do not exclude them to slow down the gainer after
the mass-transfer phase. We show that some orbital key parameters of the
system (e.g. the orbital separation) strongly depend on the spin-down
mechanism.

Our new hotspot prescription leads to a less conservative evolution than
the \citet{2008A&A...487.1129V} formalism. We showed that the opacity of
the impacted material and the geometry of the accretion (through the
$\tilde{K}$ parameter) are two highly-sensitive, badly-constrained key
parameters for the computation of non-conservative models, stressing the
need for observations and/or hydrodynamic simulations of hotspots.

A better understanding of disc formation around critically-rotating
accretors, as well as observational constraints on the magnetic-field
strength of these stars are needed. In the future, we wish to establish a
grid of non-conservative evolutions using \textsc{Binstar} to perform a
statistical study, compare with observational data and evaluate the impact
of the various spin-down mechanisms and the hotspot formalism on the
properties of the binary remnant.

\begin{acknowledgements}
  We are most grateful to Walter van Rensbergen and Jean-Pierre de
  Gr\`{e}ve for sharing their great expertise and for many fruitful
  discussions. RD and PJD acknowledge support from the Communaut\'{e}
  fran\c{c}aise de Belgique -- Actions de Recherche Concert\'{e}es. LS is
  an FNRS Researcher. PJD is Charg\'{e} de Recherche (FNRS-F.R.S.).
\end{acknowledgements}

\bibliographystyle{aa}
\bibliography{papier}

\begin{appendix}
\section{Hotspots and Critical Mass-transfer rate}
\label{ap:hotspot}

In this appendix, we derive a limit for the gainer's mass-accretion rate by
imposing that the luminosity at the hotspot location always remains below
the critical Eddington value (Eq.~(\ref{eq:Ledd})). In the case of direct
impact, the kinetic energy of the gas stream emerging from the inner
Lagrangian point $\mathcal{L}_1$ is partly converted into thermal energy
causing a hotspot, and partly radiated away.

To calculate the accretion luminosity ($L_{\mathrm{acc}}$), we estimate the
Jacobi constant (only conserved quantity in our problem) of a test particle
at $\mathcal{L}_1$. In the rotating frame
\begin{eqnarray}
  C_{\mathcal{L}_{1}} & = &
  -G \left( \frac{M_{\mathrm{d}}}{R_{\mathrm{d}}} +
    \frac{M_{\mathrm{g}}}{l_{\mathrm{g}}} \right) - \frac{1}{2}
  r_{\mathcal{L}_{1}}^{2} \Omega_{\mathrm{orb}}^{2} \, \left( +\frac{1}{2}
    v_{\mathcal{L}_{1}}^{2}\right) , 
\end{eqnarray}
where $l_{\mathrm{g}}$ is the distance between the centre of the gainer and
$\mathcal{L}_1$, $r_{\mathcal{L}_{1}}$ is the distance between
$\mathcal{L}_1$ and the center of mass of the system, $v_{\mathcal{L}_{1}}$
is the initial speed of the stream at $\mathcal{L}_1$ in the rotating
frame, and $R_{\mathrm{d}}$ is the radius of the donor filling its Roche
lobe. The last term is usually negligible because the velocity of the
particle at $\mathcal{L}_1$ (assumed to be equal to the sound speed) is
much smaller than the orbital velocity ($v_{\mathcal{L}_{1}} \ll
r_{\mathcal{L}_{1}}\Omega_{\mathrm{orb}})$. Similarly, at the impact
location,
\begin{eqnarray}
  C_{\mathrm{imp}}&=& -G\left(\frac{M_{\mathrm{d}}}{l_{\mathrm{d}}}+
    \frac{M_{\mathrm{g}}}{R_{\mathrm{g}}}\right)-\frac{1}{2}r_{\mathrm{g}}^{2}
  \Omega_{\mathrm{orb}}^{2} + \frac{1}{2}v_{\mathrm{imp}}^{2} \\
  &=& e_{\mathrm{imp}} + \frac{1}{2}v_{\mathrm{imp}}^{2} , 
\end{eqnarray}
where $l_{\mathrm{d}}$ is the distance between the centre of the donor
and the hotspot, $v_{\mathrm{imp}}$ is the stream velocity in the
rotating frame at the impact location and $r_{\mathrm{g}}$ is the
distance between the center of mass of the system and the hotspot. Since
$C_{\mathcal{L}_{1}}=C_{\mathrm{imp}}$, the energy gained by the stream
writes
\begin{equation}
  \frac{1}{2}v_{\mathrm{imp}}^{2}=C_{\mathcal{L}_{1}}-e_{\mathrm{imp}}
  \label{eq:v_stream}
\end{equation}
and is partially radiated away by the hotspot and imparted to the star in
the form of internal energy. Note that in \textsc{Binstar}, the stream
velocity at impact $v_{\mathrm{imp}}$ is consistently computed by following
the ballistic motion of the stream (see Sect.~\ref{sec:MOD:hot}) in the
rotating frame. Before the accretion starts, the gainer's total energy (in
the rotating frame, e.g. \citealt{2000taap.book.....P}) is given by
\begin{equation}
  U_{\mathrm{g}}=E_{\mathrm{kin}}+E_{\mathrm{grav}}+E_{\mathrm{i}}+
  E_{\mathrm{nuc}}+E_{\mathrm{rot}}-\frac{1}{2} M_{\mathrm{g}}
  a_{\mathrm{g}}^{2}\Omega_{\mathrm{orb}}^{2} , \label{eq:U_gainer}
\end{equation}
where $a_{\mathrm{g}}$ is the distance between the center of mass of the
system and the center of mass of the gainer, and after the mass deposition
\begin{equation}
  U_{\mathrm{g}}^{\prime} = E_{\mathrm{kin}}^{\prime} +
  E_{\mathrm{grav}}^{\prime}+E_{\mathrm{i}}^{\prime} +
  E_{\mathrm{nuc}}^{\prime} + E_{\mathrm{rot}}^{\prime} -
  \frac{1}{2}M_{\mathrm{g}}^{\prime}a_{\mathrm{g}}^{2}\Omega_{\mathrm{orb}}^{2} ,
  \label{eq:Uprime_gainer}
\end{equation}
where $E_{\mathrm{rot}}=\frac{1}{2}I_{\mathrm{g}}\Omega_{\mathrm{g}}^{2}$
and $E_{\mathrm{kin}}$ are the rotational and radial motion kinetic energy
of the star, and $E_{\mathrm{grav}}$, $E_{\mathrm{i}}$ and
$E_{\mathrm{nuc}}$ the gravitational, internal and nuclear energy content
of the star, respectively \citep{1990sse..book.....K}. Conservation of the
total energy of the system composed of the gainer and accreted material (we
neglect changes in the donor's energy) writes
\begin{equation}
  \left(U_{\mathrm{g}}^{\prime} - U_{\mathrm{g}}
  \right) +\frac{1}{2} M_{\mathrm{acc}} v_{\mathrm{imp}}^{2} + \mathrm{d}
  U_{\mathrm{rad}} + \mathrm{d}U_{\mathrm{acc}} = 0 , \label{eq:eg_conservation}
\end{equation}
where $M_{\mathrm{acc}}=\dot{M}_{\mathrm{acc}}\times\Delta t$ is the
accreted mass, $\mathrm{d}U_{\mathrm{rad}}$ the energy radiated away by the
star and $\mathrm{d}U_{\mathrm{acc}}$ the energy lost (by radiation) during
the accretion process. For a star in hydrostatic equilibrium, we have
\begin{equation}
  L = \frac{\mathrm{d}U_{\mathrm{rad}}}{\mathrm{d}t} =
  -\frac{\mathrm{d}}{\mathrm{d}t}\left(E_{\mathrm{kin}} + E_{\mathrm{grav}} +
  E_{\mathrm{i}} +
  E_{\mathrm{nuc}}\right) .\label{eq:star_eg_conservation}
\end{equation}
In the latter expression, we did not consider the rotational kinetic energy
because it is not accounted for in the stellar structure equations (there
is no feed back of rotation on the star's structure). Inserting
Eqs.~(\ref{eq:v_stream}), (\ref{eq:U_gainer}) and (\ref{eq:Uprime_gainer})
into (\ref{eq:eg_conservation}) with the help of
(\ref{eq:star_eg_conservation}) we obtain
\begin{eqnarray}
  \mathrm{d}U_{\mathrm{acc}} & = & L_{\mathrm{acc}}\times\Delta t\\ & = &
  \frac{1}{2} M_{\mathrm{acc}} v_{\mathrm{imp}}^{2} + E_{\mathrm{rot}} -
  E_{\mathrm{rot}}^{\prime} - \frac{1}{2} a_{\mathrm{g}}^{2}
  \Omega_{\mathrm{orb}}^{2} \left(M_{\mathrm{g}} -
    M_{\mathrm{g}}^{\prime} \right)\\ & = &
  \frac{1}{2} M_{\mathrm{acc}} v_{\mathrm{imp}}^{2} + E_{\mathrm{rot}} -
  E_{\mathrm{rot}}^{\prime} + \frac{1}{2} M_{\mathrm{acc}}
  a_{\mathrm{g}}^{2} \Omega_{\mathrm{orb}}^{2} .
\end{eqnarray}
In a binary system with rotating stars, the accretion luminosity finally
writes
\begin{equation}
  L_{\mathrm{acc}} = \frac{1}{\Delta t}
  \left(\frac{1}{2}M_{\mathrm{acc}}v_{\mathrm{imp}}^{2} + E_{\mathrm{rot}}
  -E_{\mathrm{rot}}^{\prime} + \frac{1}{2} M_{\mathrm{acc}}
  a_{\mathrm{g}}^{2} \Omega_{\mathrm{orb}}^{2}
  \right) . \label{eq:Eacc}
\end{equation}
To account for the fact that not all the accretion luminosity is radiated
away\footnote{Hence, the term `accretion luminosity' for the quantity
  expressed by Eq.~(\ref{eq:Eacc}) is not especially well suited!
  Unfortunately, it is accepted terminology.} (since some fraction is used
to increase the internal energy), we introduce the factor
$\alpha_{\mathrm{acc}}<1$ to evaluate the actual luminosity associated with
accretion:
\begin{equation}
\label{eq:Lacc}
L_{\mathrm{acc,rad}} = \alpha_{\mathrm{acc}} L_{\mathrm{acc}}.
\end{equation}

The matter at the hotspot location will be ejected if the hotspot
luminosity is larger than the gainer's Eddington luminosity
$L_{\mathrm{edd,g}}$:
\begin{equation}
  \label{eq:Ledd2}
  L_{\mathrm{edd,g}} = \frac{4 \pi c G M_{g}}{\kappa},
\end{equation}
where $\kappa$ is the plasma opacity. There are a few subtleties to handle
properly, however. First, for rotating stars, we should use the Eddington
luminosity corrected for the effects of stellar rotation
\citep{2009pfer.book.....M}
\begin{eqnarray}
  \label{eq:Ltilde}
  \tilde{L}_{\mathrm{edd,g}} & = &
  \left(1-\Gamma\right)L_{\mathrm{edd,g}}\\ 
  & = & \left(1-\frac{2
    \Omega_{\mathrm{spin,g}}^{2} R_{\mathrm{g}}^{3}} {3 G
    M_{\mathrm{g}}} \right) L_{\mathrm{edd,g}} .
  \label{eq:Ltilde2}
\end{eqnarray}
Second, the stellar luminosity contributes to the accretion luminosity in
powering the mass ejection from the hotspot. However, the surfaces involved
are different, and this must be taken into account, especially in the
comparison with the Eddington luminosity, whose 4$\pi$ factor in expression
Eq.~(\ref{eq:Ledd2}) implicitly assumes that the whole sphere contributes
to the luminosity. Therefore, for the purpose of this comparison, it is
necessary to multiply $L_{\mathrm{acc,rad}}$ by the factor
$S_{\mathrm{star}} / S_{\mathrm{hotspot}}$ to make all luminosities
correspond to the same surface:
\begin{equation}
  L_{\mathrm{hotspot}}= L_{\mathrm{acc,rad}}
  \times\frac{S_{\mathrm{star}}}{S_{\mathrm{hotspot}}} =
  \alpha_{\mathrm{acc}} \times L_{\mathrm{acc}}
  \times\frac{S_{\mathrm{star}}}{S_{\mathrm{hotspot}}} , 
  \label{eq:LHS}
\end{equation}
or in other words 
\begin{eqnarray}
  \label{eq:L_hotspot}
  L_{\mathrm{hotspot}} & = & \tilde{K} L_{\mathrm{acc}} \\ 
  \tilde{K} & = & \alpha_{\mathrm{acc}}
  \frac{S_{\mathrm{star}}}{S_{\mathrm{hotspot}}} ,
\end{eqnarray}
where $S_{\mathrm{hotspot}}$ and $S_{\mathrm{star}}$ are the hotspot and
gainer's surface areas, respectively. Values of $\tilde{K}$ have been
calibrated by \cite{2011A&A...528A..16V} and we use their prescription
\begin{equation}\label{eq:ktilde}
  \tilde{K} = 3.9188 \left( \frac{M_{\mathrm{d}}+M{\mathrm{g}}}{M_{\odot}}
  \right)^{1.645} ,
\end{equation}
which only depends on the total mass of the system. Thus, matter at the
hotspot location will be ejected if
\begin{equation}
  L_{\mathrm{g}} + L_{\mathrm{hotspot}} > \tilde{L}_{\mathrm{edd,g}} ,
\label{eq:Edd_criterion}
\end{equation}
where $\tilde{L}_{\mathrm{edd,g}}$ is the
gainer's Eddington luminosity, and $L_{\mathrm{g}}$ its luminosity.

The critical rate above which the hotspot luminosity prevents matter from
being accreted is given by solving for $\dot{M}_{\mathrm{acc}}$ in
Eq.~(\ref{eq:Eacc}), combined with Eqs.~(\ref{eq:L_hotspot}),
(\ref{eq:Edd_criterion}) and (\ref{eq:Lacc}), and yielding
(Eq.~(\ref{eq:Macc_crit})):
\begin{equation}
  \dot{M}_{\mathrm{acc}}^{\mathrm{crit}}={ \frac{2}{v_{\mathrm{imp}}^{2} +
      a_{\mathrm{g}}^{2} \Omega_{\mathrm{orb}}^{2}}}\left(
    \frac{\tilde{L}_{\mathrm{edd,g}}-L_{\mathrm{g}}}{\tilde{K}}
    + \frac{E_{\mathrm{rot}}^{\prime}-E_{\mathrm{rot}}}{\Delta
      t}\right) .
\end{equation}
When the mass-transfer rate exceeds the critical value, no mass can be
accreted.

\cite{2008A&A...487.1129V} derived a different expression for
$\dot{M}_{\mathrm{acc}}^{\mathrm{crit}}$ (their Eq.~(18)) which, using our
notations, writes
\begin{equation}
  \dot{M}_{\mathrm{acc}}^{\mathrm{crit}} = \frac{2}{v_{\mathrm{imp}}^{2}} \left[
    \left( \Delta E \frac{R_{\mathrm{g}}L_{\mathrm{edd},g}}{G
        M_{\mathrm{g}}}
      - L_{\mathrm{g}} \right) \frac{1}{\tilde{K}} + 
    \frac{E_{\mathrm{rot}}^{\prime} -
      E_{\mathrm{rot}}}{\Delta t} \right] ,\label{eq:VUB_macc}
\end{equation}
where
\begin{eqnarray}
  \Delta E &=& (-E_{\mathrm{grav}})+(-E_{\mathrm{orb}})+(-E_{\mathrm{eq}})\\
  &=& G\left(\frac{M_{\mathrm{d}}}{S_{\mathrm{d}}}+\frac{M_{\mathrm{g}}}
    {R_{\mathrm{g}}}\right) + \frac{1}{2} r^{2} \Omega_{\mathrm{orb}}^{2} 
  - \frac{1}{2} R_{\mathrm{g}}^{2}\Omega_{\mathrm{spin,g}}^{2} .
\end{eqnarray}
Based on our simulations of a 6 + 3.6~M$_{\odot}$ system with initial
period P$_{\mathrm{init}}$ = 2.5~days, we find that $E_{\mathrm{grav}}
\approx 2-3 \times |E_{\mathrm{eq}}| \gg E_{\mathrm{orb}}$ which allows
us to simplify Eq.~(\ref{eq:VUB_macc}) (neglecting the $E_{\mathrm{rot}}$
terms)

\begin{eqnarray}
  \dot{M}_{\mathrm{acc}}^{\mathrm{crit}} 
  & \approx &
  \frac{2}{v_{\mathrm{imp}}^{2}\tilde{K}} \left[ \left( 1 +
      \frac{E_{\mathrm{eq}}}{E_{\mathrm{pot}}} \right)
    L_{\mathrm{edd,g}} - L_{\mathrm{g}} \right]\\
  & \approx & \
  \frac{2}{v_{\mathrm{imp}}^{2} \tilde{K} } \left[ \left( 1 - \frac{
        \Omega_{\mathrm{spin,g}}^{2} R_{\mathrm{g}}^{3} }
      { 2 G M_{\mathrm{g}}} \right) L_{\mathrm{edd,g}} -
    L_{\mathrm{g}} 
  \right] \\
  & \approx &
  {\frac{2}{v_{\mathrm{imp}}^{2}\tilde{K} }
  } \left( \breve{L}_{\mathrm{edd,g}} - L_{\mathrm{g}} \right) .
\end{eqnarray}
This formulation is very comparable to our Eq.~(\ref{eq:Macc_crit}), in
case $E_{\mathrm{rot}}$ is small, except for the term $a_{\mathrm{g}}^{2}
\Omega_{\mathrm{orb}}^{2}$ in the denominator and for some numerical
factors of order unity.

\end{appendix}

\end{document}